\documentclass[12pt,preprint]{aastex}

\newcommand{\msb}{$M_{\odot}$~}
\newcommand{\ms}{$M_{\odot}$}

\begin{document}

\title{Galactic evolution of Sr, Y, Zr: \\
a multiplicity of nucleosynthetic processes.}

\author{Claudia Travaglio\altaffilmark{1,2}}
\affil{1. Max-Planck Institut f\"ur Astrophysik, Karl-Schwarzschild Strasse 1,
D-85741 Garching bei M\"unchen, Germany}

\affil{2. Istituto Nazionale di Astrofisica (INAF) - Osservatorio Astronomico 
di Torino, via Osservatorio 20, 10025 Pino Torinese (TO), Italy}
 
\author{Roberto
 Gallino\altaffilmark{3}, Enrico Arnone\altaffilmark{3,4}}
\affil{3. Dipartimento di Fisica Generale, Universit\'a di Torino, and Sezione 
INFN di Torino, Via P.Giuria 1, I-10125 Torino, Italy}

\affil{4. Department of Physics and Astronomy, The Open University, Walton Hall
Milton Keynes, MK76AA, UK} 

\author{John Cowan, Faith Jordan\altaffilmark{5}}
\affil{5. Department of Physics and Astronomy, University of Oklahoma, 
Norman, OK 73019}

\author{Christopher Sneden\altaffilmark{6}}
\affil{6. Department of Astronomy and McDonald Observatory, University of 
Texas, Austin, TX 78712}

\begin{abstract}

In this paper we follow the Galactic enrichment of three easily observed 
light $n$-capture elements --  Sr, Y, and Zr. Input stellar yields have 
been first separated into their respective main and weak $s$-process components, 
and $r$-process component. The $s$-process 
yields from Asymptotic Giant Branch (AGB) stars of low to intermediate mass 
are computed, exploring a wide range of efficiencies of the major neutron source, 
$^{13}$C, and covering both disk and halo metallicities. AGB stars have been 
shown to reproduce the main $s$-component in the solar system, i.e., the 
$s$-process isotopic distribution of all heavy isotopes with atomic mass number 
A $>$ 90, with a minor contribution to the light $s$-process isotopes up to 
A $\sim$ 90. The concurrent weak $s$-process, which accounts for the major fraction
of the light $s$-process isotopes in the solar system and occurs in massive stars 
by the operation of the $^{22}$Ne neutron source, is discussed in detail.
Neither  the main $s$-, nor the weak $s$-components are shown to contribute significantly
to the neutron capture element abundances observed in unevolved 
halo stars.  Knowing the $s$-process distribution at the epoch of the solar 
system formation, we first employed  the $r$-process residuals method to infer the 
isotopic distribution of the $r$-process. We assumed a primary $r$-process 
production in the Galaxy from moderately massive Type II supernovae that best 
reproduces the observational Galactic trend of  metallicity versus  Eu, an almost 
pure $r$-process element. We present a detailed analysis of a large published 
database of spectroscopic observations of Sr, Y, Zr, Ba, and Eu for Galactic 
stars at various metallicities, showing that the observed trends versus 
metallicity can be understood in light of a multiplicity of stellar 
neutron-capture components. Spectroscopic observations of the Sr, Y, and Zr to 
Ba and Eu abundance ratios versus metallicity provide  useful diagnostics of the 
types of neutron-capture processes forming Sr, Y and Zr. In particular, the 
observed [Sr,Y,Zr/Ba,Eu] ratio is clearly not flat at low metallicities, as we would 
expect if Ba, Eu and Sr, Y, Zr all had the same $r$-process nucleosynthetic 
origin. We discuss our chemical evolution predictions,  taking into account the 
interplay between different processes to produce Sr-Y-Zr.  
Making use of the very $r$-process-rich and very metal-poor stars like CS~22892-052 
and CS~31082-001, we find hints, and discuss the possibility of 
a {\it primary process} in low-metallicity massive stars, different from the 
`classical $s$-process' and from the `classical $r$-process', that we tentatively define 
LEPP (Lighter Element Primary Process). This allows us to 
revise the estimates of the $r$-process contributions to the solar Sr, Y and Zr abundances, 
as well as of the contribution to the $s$-only isotopes $^{86,87}$Sr and $^{96}$Mo.

\end{abstract}


\section{Introduction}

In order to reconstruct the solar system composition of the heavy elements
beyond Fe, two major neutron capture mechanisms have been invoked 
since the classical work by Burbidge et al.~(1957): the slow ($s$) process 
and the rapid ($r$) process.
The $s$-process path requires a relatively low neutron density, 
$n_n$ $<$ 10$^8$ cm$^{-3}$, and moves along the valley of $\beta$ stability. 
This builds up approximately 
half the nuclides from Fe to Bi, in particular feeding the 
elements Sr-Y-Zr, Ba-La-Ce-Pr-Nd, and Pb, which define  the three 
major abundance $s$-peaks. 
The sources for the required free neutrons can be either the reaction
$^{22}$Ne($\alpha$,~n)$^{25}$Mg or $^{13}$C($\alpha$,~n)$^{16}$O.

Since the first phenomenological analysis, the so-called {\it classical} 
analysis (Clayton et al.~1961; Seeger et al.~1965), 
the $s$-process abundance distribution in the solar system has been 
recognized as arising from a non-unique site.  
At least three components have been required: the {\it main}, the {\it weak}, 
and the {\it strong} $s$-component (Clayton \& Ward 1974; K\"appeler et
al.~1982; K\"appeler, Beer, \& Wisshak~1989).

The {\it main} $s$-component, accounting for the $s$-process isotopic
distribution in the atomic mass number range  90~$<$~A~$<$~208, was shown 
to occur in low-mass ($M \lesssim$ 4 \ms) Asymptotic Giant Branch stars (hereafter
AGB) during recurrent thermal instabilities developing above the He-burning 
shell (see Busso, Gallino, \& Wasserburg 1999 for a review). 
The whole He intershell, that is the region comprised
between the H-shell and the He-shell, becomes convective for a short period
of time (the convective thermal pulse, hereafter TP). 
During the AGB phase, after the quenching of a TP, the convective envelope
penetrates below the H-He discontinuity  ({\it third dredge-up} episode,
hereafter TDU), mixing to the surface freshly synthesized $^{4}$He, $^{12}$C 
and $s$-process elements. The maximum temperature in the deepest region of 
the convective TP barely 
reaches $T$ = 3 $\times$ 10$^8$ K; at this temperature the $^{22}$Ne neutron 
source is marginally activated, and the $^{13}$C source plays the major role 
for the main $s$-component. At TDU,  the H-rich envelope and the
He intershell coming  into contact favors the penetration of a small amount of
protons into the top layers of the He- and C-rich zones. 
At hydrogen 
re-ignition, protons are captured by the abundant $^{12}$C, giving rise to the 
formation of a so-called $^{13}$C {\it pocket}. Stellar model calculations for 
the AGB phases by Straniero et al.~(1995, 1997) showed that all the 
$^{13}$C nuclei 
present in the $^{13}$C-pocket are consumed locally in the 
radiative layers of the 
He intershell, before a new TP develops. This provides an $s$-process 
abundance distribution that is strongly dependent on the initial 
metallicity (Gallino et al.~1998; Busso et al. 2001).

The {\it weak} $s$-component, responsible for a major contribution to
the $s$-process nuclides up to A $\simeq$ 90, has been recognized as the result 
of neutron capture synthesis in advanced evolutionary phases of massive stars. 
Previous studies (Lamb et al.~1977; Arnett \& Thielemann~1985; Prantzos et 
al.~1990; Raiteri et al.~1991a; The, El Eid, \& Meyer~2000) have concentrated 
on the reaction $^{22}$Ne($\alpha$,~n)$^{25}$Mg as the major neutron source for 
this process. The $^{22}$Ne neutron source is activated partly in the 
convective core He-burning 
and partly in the subsequent convective C-burning shell phase (Raiteri et al.
1991b; 1993). The $s$-process in massive stars is metallicity dependent, since 
$^{22}$Ne is produced 
 from the conversion of CNO nuclei into $^{14}$N in the H-burning
shell followed by double $\alpha$-capture on $^{14}$N in the early phases of
He burning (Prantzos et al. 1990; Raiteri et al. 1992). As we will discuss in \S~5.3, 
additional neutron sources, partly of primary origin, may take place in the inner 
regions of massive stars during convective shell C-burning (Arnett \& Truran~1969;
Thielemann \& Arnett~1985; Arnett \& Thielemann~1985; Raiteri et al.~1991b) and, 
more importantly, during explosive nucleosynthesis in the O-rich regions (Hoffmann, 
Woosley, \& Weaver~2001; Heger et al.~2001; Heger \& Woosley~2002; Rauscher et 
al.~2002; Woosley, Heger, \& Weaver~2002; Limongi \& Chieffi~2003).

Finally, the {\it strong} $s$-component was introduced by Clayton \&
Rassbach~(1967) in order to reproduce more than 50\% of solar $^{208}$Pb,
the most abundant Pb isotope.
Recent studies by Gallino et al.~(1998) and Travaglio et al.~(2001a)
demonstrated that the role attributed to the strong $s$-component is played
by low-metallicity ([Fe/H] \footnote{In this paper we follow the
usual convention of identifying overall metallicity with the stellar
[Fe/H] value, following the standard notation that
[X/Y]~$\equiv$ log$_{\rm 10}$(N$_{\rm X}$/N$_{\rm Y}$)$_{\rm star}$~--
log$_{\rm 10}$(N$_{\rm X}$/N$_{\rm Y}$)$_{\odot}$}
$< -$1.5) low mass AGB stars.

The $r$-process, however,
 takes place in an extremely neutron-rich environment 
in which the mean time between successive neutron captures is very short 
compared with the time to undergo a $\beta$-decay.
Supernovae are currently believed to be the site 
 of the $r$-process. However, there 
have been many attempts to define the right physical conditions 
for the $r$-process to occur (e.g., Hillebrandt~1978; Mathews \& Cowan~1990; 
Woosley et al.~1994; Wheeler, Cowan, \& Hillebrandt~1998).
Three possible sites have been discussed in recent works.
The  first possibility relies on neutrino-powered winds of a young neutron star 
(Duncan, Shapiro, \& Wasserman~1986; Woosley et al.~1994; Takahashi, 
Witti, \& Janka~1994). Recently, Thompson, Burrows, \& Meyer~(2001) argued 
that it may be difficult to achieve the necessary high entropy and short 
timescales in the ejecta in order to reproduce the solar system $r$-process 
abundance distribution.
A second possibility is related to the merging of two neutron stars in a
binary system and has been examined by Freiburghaus, Rosswog, \& Thielemann~(1999). 
However, Qian~(2000) argued that the predicted amount of $r$-process ejecta in
metal-poor stars would be too high in $r$-elements with $A <$ 130 and with 
$A >$ 130 as compared with spectroscopic abundances of metal-poor stars 
and that  the event rate would be too low. The third possibility 
relies upon asymmetric explosions of massive stars and jet-like outflows
in the nascent neutron star (LeBlanc \& Wilson~1970; Cameron~2001,~2003). 
Each of these proposed sites faces major problems, including 
reaching the required physical conditions without ad hoc assumptions to
produce a satisfactory fit to the solar system $r$-process pattern. 
Hence, the stellar source for $r$-process abundances is still a matter of 
debate.  
Moreover, it has been suggested that at least two different supernova 
sources are required for the synthesis of $r$-process nuclei below and 
beyond the neutron magic number N = 82 (Wasserburg, Busso, \& Gallino~1996; 
Sneden et al.~2000a).

Neutron-capture elements observed in Pop.~II field stars are generally
interpreted in an observational framework developed more than 20 years ago.
Spite \& Spite~(1978) first demonstrated that observations of Ba 
(a predominantly, 80\%,  $s$-process element in
solar system material) 
and Eu (an $r$-element, 95\%,  in the solar system)
exhibit a non-solar abundance 
pattern in unevolved halo stars, with [Eu/Ba] $>$ 0. This was 
interpreted by Truran~(1981) as evidence of an $r$-process 
nucleosynthesis signature at low metallicities, with little 
evidence for $s$-process contributions. Observational support for this 
view has grown both in large-sample surveys (e.g., Gilroy et
al. 1988; McWilliam et al. 1995) and in detailed analyses of several 
ultra-metal-poor ([Fe/H]~$\lesssim$ $-$2.5) $r$-process-rich stars
(e.g., Sneden et al. 2000a; Westin et al. 2000; Hill et al. 2002).
The abundances of the heavier $n$-capture elements (Z~$\geq$~56)
in such stars often is an excellent match to a scaled solar-system
$r$-process distribution (e.g., Cowan et al. 2002), but $n$-capture
abundances of lighter elements below Ba often show significant departures 
from this distribution.
It is not obvious how the observed abundances of the lighter $n$-capture
elements in metal-poor stars evolve to those seen in the solar system and
in Pop. I stars.

A detailed $r$- and $s$-process decomposition can be obtained for the
solar system, based on the experimental knowledge of neutron capture cross 
sections and on the isotopic analysis of meteoritic samples that best represent 
the protosolar nebula composition. Unfortunately, the solar system composition 
only provides a single data point in the time evolution of $n$-capture 
elements in the Galaxy. Investigations into the chemical composition of 
matter at different epochs can only be accomplished through high-resolution 
stellar spectroscopic abundance studies. Although the correlation of metallicity 
with time is hardly perfect, stars with sub-solar metallicities are tracers of 
the chemical compositions of the gas at different times of evolution of the Galaxy. 
Elemental $n$-capture abundances of field Galactic stars at different 
metallicities show two main characteristics: first, an average trend 
to lower [X/Fe] with decreasing [Fe/H]; second, a dispersion in [X/Fe] that 
increases with decreasing [Fe/H]. 
Theorists have argued that the large dispersions arise from local chemical 
inhomogeneities in the interstellar medium of heavy elements 
(in particular Ba, Eu and Sr), due to incomplete mixing of the gas 
in the Galactic halo (see Tsujimoto, Shigeyama, \& Yoshii~1999; 
Ikuta \& Arimoto~1999; Raiteri et al.~1999; Argast et al.~2000,~2002; 
Travaglio, Galli, \& Burkert~2001b). 

Spite \& Spite~(1978) were the first to find observational evidence of a 
trend of declining [Ba/Fe] and [Y/Fe] below [Fe/H] $\sim$ $-$2. 
Unfortunately their sample of 11 stars was too small to find those rarer 
stars with super-solar $n$-capture abundances, or to detect 
the intrinsic dispersion in these ratios. 
The earliest evidence for a dispersion at low metallicity came from 
Griffin et al.~(1982), who found very strong Eu lines in the halo star 
HD~115444 ([Fe/H] $\sim$ $-$3), subsequently confirmed by the studies of 
Gilroy et al.~(1988), Sneden et al.~(1998) and  Westin et 
al.~(2000). Other similar well known examples are stars extremely rich
in $r$-process elements with respect to a solar-scaled composition 
at the observed [Fe/H], like CS~22892-052 (Sneden et al.~2000a, 2003a 
and references therein), CS~31082-001 (Cayrel et al.~2001; Hill et
al.~2002), or stars of comparable metallicity  but showing a much lower
$n$-capture element enhancement, like
HD~122563 (Westin et al.~2000). Other well observed stars are
BD~+17~3248 (Cowan et al.~2002) and CS~22949-037 (Depagne et al.~2002). 
Studies by  Gilroy et al.~(1988), Ryan et al.~(1991, 1996), 
Gratton \& Sneden~(1988, 1994), McWilliam et al.~(1995), McWilliam~(1998), 
and more recently Burris et al.~(2000), Fulbright~(2000) and Johnson \& 
Bolte~(2002) have found dispersions in $n$-capture elements/Fe ratios of more 
than a factor of 100 from star to star at a given metallicity. 

In this paper we study the Galactic chemical evolution (hereafter GCE) of 
Sr, Y, and Zr. The paper is organized as follows: in \S~2 we focus on the Sr-Y-Zr 
production by AGB stars at different metallicities. 
In \S~3 we briefly review the GCE model adopted and our $r$-process 
assumptions. In \S~4 we present our collection of spectroscopic abundances 
in field stars at different metallicities, updated from the recent 
literature. The unique compositions of some of the stars of our sample will 
be discussed. In \S~5 we first discuss how the main $s$-process 
nucleosynthesis in AGB stars reproduces a major fraction of
the solar isotopic compositions of Sr-Y-Zr 
by following their enrichment throughout the Galactic history. 
We then examine the minor role played by the weak $s$-process in massive stars 
to the solar system inventory of the first $s$-peak abundances. Both
the main and the weak $s$-process do not affect the heavy element abundances
of unevolved stars at low metallicities. However, we  discuss 
the complex nature of neutron captures occurring in advanced stages of
massive stars and the possibility of activation of primary neutron sources during
shell C-burning or explosive nucleosynthesis in the oxygen-rich regions,
not related to the classical $s$- or $r$-process.
A general comparison of our $s$-process predictions is then made with spectroscopic 
observations of Sr, Y, Zr in field stars at different metallicities. In particular, 
we make use of the spectroscopic observations in extremely
$r$-process-rich and very metal-poor stars, like CS 22892-052, to infer
the $r$-process fraction of Sr, Y, Zr that is strictly related to the main
$r$-process feeding the heavy elements beyond Ba. 
We also examine very recent spectroscopic observations of heavy elements
in dwarf spheroidal galaxies (Shetrone et al.~2001; Shetrone et al.~2003;
Tolstoy et al.~2003), as well as in the globular cluster M15 (Sneden et al.~2000b).
In particular we investigate how they compare with the Galactic trend versus 
metallicity. 
In \S~6 we show how a extra primary process (not yet fully quantified 
from the present status of nucleosynthesis models) is needed to fully explain 
the solar composition of Sr-Y-Zr and in particular their Galactic trend at 
very low metallicities. 
Finally, in \S~7 we summarize the main conclusions of this work and 
point out several areas that deserve further analysis. 

\section{Sr-Y-Zr production by AGB stars at different metallicities}

We modeled the AGB nucleosynthesis, as in Straniero 
et al.~(1997) and Gallino et al.~(1998), with post-process calculations
that make use of stellar evolutionary models with the FRANEC code (Frascati
Raphson-Newton evolutionary code; see Chieffi \& Straniero~1989). 
We computed stellar yields for $s$-process elements in AGB stars 
injected in the interstellar medium by mass loss winds from stars of
different masses. This has been done for both Galactic halo and disk 
metallicities and for a wide range of $^{13}$C pocket efficiencies (see
Travaglio et al.~1999, 2001a for applications of these techniques
to heavier $n$-capture elements, with Z~$\geq$~56). 
The cumulative He intershell mass dredged up into the envelope and ejected 
into the interstellar medium is reported in Travaglio et al.~(2001a, 
their Table~1). 

In spite of the fact that successful models for the formation of the 
$^{13}$C-pocket have been advanced (Hollowell \& Iben~1988; Herwig et al.~1997; 
Langer et al.~1999; Cristallo et al.~2001), the mass involved and the profile of 
the $^{13}$C-pocket must  still be considered as free parameters, given the 
difficulty of a sophisticated treatment of the hydrodynamical behavior at the H/He 
discontinuity during each TDU episode.
However, a series of constraints have been obtained by comparing spectroscopic 
abundances in $s$-enriched stars (MS, S, C, post-AGB, Ba and CH stars)
at different metallicities with AGB model predictions (see e.g., Busso et 
al.~1995,~2001; Abia et al.~2001,~2002 and references therein). 
These authors conclude that the observations in general confirm the complex 
dependence of neutron captures on metallicity. The spread observed in the abundance 
ratio of the Ba-peak elements with respect to the lighter Zr-peak elements 
requires the existence of an intrinsic spread in the mass of the $^{13}$C-pocket. 
The same spread in the $^{13}$C concentration has been found to 
be appropriate to match the $s$-process isotopic signature of heavy elements 
in presolar grains condensed in circumstellar envelopes of AGB stars (see e.g., 
Lugaro et al.~1999,~2003 and references therein). 

In our calculations, the intrinsic spread in the $s$-process yields at each 
metallicity has been modeled parametrically by varying the $^{13}$C 
concentration in the pocket from 0 up to a factor of 1.5$\times$ the `standard' 
value of $\sim$4 $\times$ 10$^{-6}$ \msb of $^{13}$C (Gallino et 
al.~1998; their ST case). Note that a maximum abundance of $^{13}$C
is expected in the pocket. Indeed, too high a concentration of protons, that may 
diffuse into the top layers of the pocket, would activate the concurring reaction 
$^{13}$C(p,~$\gamma$)$^{14}$N (with $^{14}$N acting as a strong neutron poison).

We show in the {\it upper panel} of Fig.~1 the elemental production factor of 
Sr, Y, Zr versus [Fe/H] obtained for AGB stars of initial 
1.5 \msb and different metallicities, for the ST choice of the $^{13}$C pocket. 
For comparison we also plot in the lower panel the 
elemental production factor of Ba, Hf (an element that also receives a major
contribution from the $s$-process), and Pb. In Fig.~2 we show the 
isotopic production factors of 
Sr ($^{86,87,88}$Sr, {\it upper panel}), Y ($^{89}$Y, {\it central panel}), 
and Zr ($^{90, 91, 92, 94, 96}$Zr, {\it lower panel}), versus [Fe/H]. 
Similar trends are obtained for the 3 \msb case. 
From this figure it is clear that the $s$-process production in AGB stars, 
driven by the primary $^{13}$C neutron source, gives rise to a wide spectrum 
of different abundance distributions that are strongly dependent on metallicity.
At solar metallicity, AGB stars produce copious amounts of $s$-process 
elements belonging to the Sr-Y-Zr peak at the neutron magic number N $=$ 50. 
For decreasing [Fe/H], more neutron per Fe seeds are available, thus bypassing 
the bottleneck at N $=$ 50 and progressively feeding elements to the second 
neutron magic peak (Ba-La-Ce-Nd) at N = 82, with a maximum production yield at 
[Fe/H] $\sim$ $-$0.6 (see e.g., the $^{138}$Ba trend of Fig.~1, and 
Travaglio et al.~1999 for more details). 
At lower metallicities there are enough neutrons per Fe seed to produce
large Pb excesses, in particular the most abundant isotope $^{208}$Pb at 
N $=$ 126 (Gallino et al.~1998; Travaglio et al.~2001a). 
The maximum Pb production occurs at [Fe/H] $=$ $-$1 
and then decreases following the decrease of the initial Fe concentration.
Note that at very low metallicities there is an important primary contribution
to $^{22}$Ne. Indeed, the progressive erosion of the bottom of the envelope 
by the H-burning shell makes some fresh $^{12}$C, mixed with the envelope by
previous TDU episodes, and  converted into $^{14}$N
and subsequently into $^{22}$Ne by double $\alpha$-capture during the 
early development of the next TP. 
The neutrons released by the $^{13}$C source in the pocket and by 
the $^{22}$Ne source in the TP are captured by the primary $^{22}$Ne and their 
progenies $^{25,26}$Mg, which act contemporaneously as neutron poisons and as seeds
for the production of the heavy $s$-elements (see discussion in Busso et al.
1999). As a consequence, at very low metallicities the Pb yield levels off
instead of further decreasing. This complex behavior of $s$-process production is 
extremely important in a GCE study.

We also considered AGB stars of intermediate mass (hereafter IMS), 
basing our analysis on stellar evolutionary models of 5 \msb and 7 \ms, 
and extrapolating over the whole metallicity and mass (5 $-$ 8 \ms) range. 
Travaglio et al.~(1999, 2001a) discussed the role of IMS stars in Ba to Eu and Pb 
production, concluding that they play a minor role in their Galactic enrichment. 
Typically in these stars the temperature at the base of the convective 
envelope reaches $T$ = 3.5 $\times$ 10$^{8}$ K, allowing $^{22}$Ne to
release neutrons efficiently via the channel $^{22}$Ne($\alpha$,~n)$^{25}$Mg 
(Iben~1975; Truran \& Iben~1977). 
The resulting high peak neutron density ($n_n >$ 10$^{11}$ cm$^{-3}$) gives rise 
to an overproduction of a few neutron-rich isotopes, such as 
$^{86}$Kr, $^{87}$Rb, and $^{96}$Zr, involved in branchings along the 
$s$-process path and particularly sensitive to the neutron density. 
The role of IMS in the light $s$-elements production will be discussed in \S~5.

In Fig.~3 we show the production yields for $^{88}$Sr  obtained
for the various assumed $^{13}$C pocket efficiencies. 
Similar trends hold for Y and Zr isotopes.
In Fig.~3, we also show with a thick line the unweighted average over this spread, 
the choice adopted for our GCE calculations. 
With the same technique Travaglio et al.~(1999) studied the GCE of the 
elements from Ba to Eu, and Travaglio et al.~(2001a) studied the role of 
AGB stars in the GCE of Pb and Bi. 
In \S~3 we update the results presented in the previous papers.

\section{Galactic evolution of Sr-Y-Zr: our 'tools'}

Besides the $s$-process yields from AGB stars described in \S~2,
we need other important tools to discuss 
the enrichment of Sr-Y-Zr in the Galaxy. 
First, we outline in \S~3.1 the main characteristics of
the GCE model that we follow. 
Since no $r$-process yields are currently available from stellar model 
calculations, we have introduced them into the GCE model 
(\S~3.2) under the simple hypothesis of a primary production from Type~II 
supernovae.

\subsection{The GCE model}

The GCE code for this work has been described in detail by 
Ferrini et al.~(1992).
The same model was adopted by Galli et al.~(1995) and  Travaglio et al. (2001c) 
to study the evolution of the light elements D, $^3$He and $^7$Li, and 
by Travaglio et al.~(1999) and Travaglio et al.~(2001a) to study the 
evolution of the heavy elements from Ba to Pb (see those papers for 
detailed descriptions, here we only briefly review the  basic features). 
The Galaxy is divided into three zones: halo, thick-disk, and thin-disk, 
whose fraction of stars, gas (atomic and molecular) and stellar remnants, 
is computed as a function of time up to the present epoch.
Stars are born with the same chemical composition of the gas from which 
they form.  
The thin-disk is formed from material falling in from the thick-disk and 
the halo.  
The star formation rate in the three zones is not assumed {\it a priori}, 
but it is obtained as the result of self-regulating processes occurring in 
the molecular gas phase, either spontaneous or stimulated by the presence 
of other stars. 
Stellar nucleosynthesis yields are treated according to the matrix formalism 
of Talbot \& Arnett~(1973). 
The halo phase lasts approximately up to [Fe/H] $\lesssim -$1.5; the 
thick-disk phase covers the interval $-$2.5 $\lesssim$ [Fe/H] $\lesssim -$1; 
the thin-disk phase starts at approximately [Fe/H] $\gtrsim -$1.5.

\subsection{The $r$-process assumption}

In halo stars the contribution of AGB stars is too low by far to account 
for the observed heavy element abundances. This is essentially due to the long
lifetime that low mass stars spend before reaching the AGB. Moreover, as we
discussed in \S~2, AGB stars of halo population preferentially produce
$s$-isotopes in  the Pb peak, with marginal production of elements in the Zr and Ba
$s$-peaks.

The almost pure $r$-process signature in halo stars was anticipated by Truran~(1981) 
on theoretical arguments, and by Spite \& Spite (1978) on observational grounds. 
For elements from Ba to Pb, our estimate of $r$-process contributions at 
$t$ = $t_\odot$ has been derived by subtracting the $s$-fractions from the solar 
abundances (i.e., the $r$-process residuals method). 
The $r$-process contributions are  treated as arising from a  
primary mechanism occurring in a subset of Type II SNe, i.e., those in the mass 
range 8 $-$ 10 $M_\odot$ (see Wheeler, Cowan, \& Hillebrandt 1998). 
Concerning the $r$-process contribution to elements lighter than barium,
and in particular to strontium, yttrium and zirconium, a more complex treatment
is needed, as will be discussed in \S~5.

\section{Sr-Y-Zr abundances in `unevolved' stars}

In order to compare our model results with observed abundance trends, 
we selected a sample of spectroscopic observations of Galactic field stars 
(mostly F and G dwarfs, and giants not obviously enriched by local
$s$-process production events) at different metallicities, updated with 
the most recent data available in literature. 
In Fig.~4 we show the data from these surveys in the usual manner, as
[Sr/Fe], [Y/Fe], and [Zr/Fe] values in the top, middle, and bottom panels,
plotted versus [Fe/H]. 
We include results from Spite \& Spite~(1978); Edvardsson et al.~(1993); 
Gratton \& Sneden~(1994); McWilliam et al.~(1995); McWilliam~(1998); 
Jehin et al.~(1999); Tomkin \& Lambert~(1999); Burris et al.~(2000); 
Fulbright~(2000); Sneden et al.~(2000a); Westin et al.~(2000); 
Norris, Ryan, \& Beers~(2001); Mashonkina \& Gehren~(2001);
Mishenina \& Kovtyukh~(2001); Hill et al.~(2002); Cowan et al.~(2002); 
Depagne et al.~(2002).
We have also included recent observations in dwarf spheroidal galaxies 
(Shetrone et al.~2001; Shetrone et al.~2003), as well as in three giants stars 
observed in the globular cluster M15 (Sneden et al.~2000b).

In Fig.~5 we show for comparison our collection of data for [Ba/Fe] (top
panel), [Eu/Fe] (middle panel, and [Ba/Eu] (lower panel) versus [Fe/H], 
over-imposed with the predicted average Galactic trend for the halo, thick
disk, and thin disk according to Travaglio et al. (1999). At disk metallicities 
the model predictions from AGB stars only are also shown for comparison. 

Before applying the same GCE model to the Sr, Y and Zr data, 
we pause to alert the reader to some observational uncertainties and limitations.  
First, the spectral lines all of these elements used in analyses of 
metal-poor field stars have been subject to extensive laboratory analyses.
Therefore, their transition probabilities have small uncertainties.
Second, almost all Sr, Y, and Zr abundance results are from ionized 
transitions of these elements. Thus,  abundance inter-comparisons among these 
elements, or comparisons with abundances of rare-earth elements (also 
derived from ionized transitions) have, in most cases, little dependence on 
choices of effective temperatures or gravities in the various studies 
included here.  
But each of these elements has some particular problems that should be 
kept in mind.

Strontium abundances are based almost exclusively on the very strong 
\ion{Sr}{2} 4077, 4215~\AA\ resonance lines.
Abundances derived from these transitions are very sensitive to the choice 
of (in particular) the microturbulent velocity parameter, which can 
vary from analysis to analysis for the same star.  
Additionally, these lines suffer some blending from other atomic and 
molecular transitions, and abundance analyses really ought to be done 
with synthetic spectrum computations rather than with the more common 
single-line equivalent width computations.
Fortunately in the rare spectra that have detectable (much weaker) 
higher-excitation \ion{Sr}{2} lines and the \ion{Sr}{1} 4607~\AA\ line, 
the abundances derived from all Sr features are in reasonable agreement
(e.g., Sneden et al.~2003a), confirming the reliability of the \ion{Sr}{2} 
resonance lines, given care in their analyses.

Second, yttrium and zirconium abundances usually are derived from a few 
lines of the ionized species of these elements.
None of these lines is usually strong enough for microturbulent velocity 
uncertainties to be an important source of abundance uncertainty.
This comes at a price: such lines of moderate strength in stars of moderate
metal-deficiency weaken to undetectable levels in the most metal-poor stars,
and thus Y and Zr abundances cannot be used to constrain GCE models below
[Fe/H]~$\sim$ --3.

Abundances of these three elements have been reported for globular 
cluster giant stars. 
However, comparisons between field and cluster abundances should be done
cautiously, because the vast majority of field star abundances are obtained
from blue-UV region spectra (where the stronger lines are), while cluster 
star spectra abundances are based on lines in yellow-red spectral region 
(where the fluxes are largest in these faint stars).
Unfortunately, \ion{Sr}{2} lines are few and very weak in the yellow-red,
resulting in few Sr abundance determinations in clusters.
Furthermore, the results for Y and Zr are so spotty that for the present 
study we have chosen not to make a detailed comparison with globular
clusters. Exceptions are three giant stars in M15 (see discussion below).

However, a few general remarks about heavier $n$-capture elemental
abundances in globular clusters are appropriate here.
First, note that the cluster metallicity range is [Fe/H]~$\gtrsim$ --2.4,
i.e, there are no ultra-metal-poor Galactic globular clusters.
In this metallicity regime the clusters have [Eu/Fe]~$\approx$ +0.4 
($\sigma$=~0.1), mostly with little star-to-star variations within 
individual clusters and very small cluster-to-cluster differences (see 
the review of Sneden et al.~2003b, and references therein).  
This overall Eu enhancement is comparable to the general level in 
field stars of the same metallicities.
Abundances of Ba and La in clusters exhibit more intra- and inter-cluster
variation, with an unweighted mean value [Ba,La/Fe]~$\approx$ +0.1 
($\sigma$=~0.2).
Combining this ratio with the mean Eu abundance discussed above yields
[Eu/Ba,La]~$\sim$ +0.3 for almost all well-observed globular clusters. 
This in turn suggests strongly that the heavier $n$-capture elements in 
globular clusters have been synthesized more heavily by the $r$-process 
than by the $s$-process, compared with the solar system composition of these 
elements.  
Therefore, in general terms, the $n$-capture elements in globular clusters 
were formed from the same enrichment episodes as was the halo field.
There are a few important exceptions to these general statements about
globular cluster $n$-capture elements.
For example, in M4, a globular cluster with metallicity [Fe/H] = $-$1.2, 
the $s$-process elements Ba and La are enhanced 
significantly beyond their nominal levels, e.g., [Eu/Ba,La]~$\sim$ 0.0
(Ivans et al. 1999).
This is very clear evidence for the extra presence of the products of
AGB star $s$-process nucleosynthesis, but why this has occurred in M4 
and not in most other clusters (e.g., not in the similar-metallicity
cluster M5; Ivans et al. 2001) is not well understood.
And in M15 (one of the most metal-poor clusters, [Fe/H]~$\approx$ --2.4),
a star-to-star scatter is found with a range of 0.5~dex, well beyond
observational uncertainties (Sneden et al. 1997, Johnson \& Bolte~2002).

Only a few stars of globular clusters have been subjected to a very detailed
$n$-capture abundance analysis. As one example, Sneden et al.~(2000b) 
obtained blue-region spectra of three giants in M15. Their representative 
points are shown in the various Figures as big open triangles.
The relative $n$-capture abundances from this study showed good agreement with 
abundances of these elements in so-called $r$-process-rich field stars,
i.e., $<$[Eu/Fe]$>$ ~$\approx$ +0.80 and $<$[Ba/Eu]$>$ ~$\approx$ --0.85.  
The [Eu/Fe] values indicate enhanced $n$-capture elemental abundances 
in these low metallicity giants, while the [Ba/Eu] ratios
are consistent with an $r$-process solar system abundance value. 
Both sets of elemental abundance ratios in the globular giants support 
other studies, based upon field halo giants, that demonstrate first the 
early onset of the $r$-process in the Galaxy and second the dominance of 
the $r$-process (as opposed to the $s$-process) in the early Galactic 
synthesis of $n$-capture elements. 
Yttrium and zirconium (but not strontium) abundances were also derived
in these stars, and the mean values were $<$[Y/Fe]$>$~$\approx$ --0.30
and $<$[Zr/Fe]$>$~$\approx$ +0.40. 
These values, being both higher and lower than solar, are suggestive of 
different synthesis histories for Y and Zr, at least, than for the heavier 
$n$-capture elements, i.e., Ba and Eu. 

At typical Galactic disk metallicities, the largest data set for Zr and Y 
abundances in field stars is that of Edvardsson et al.~(1993). The abundances from 
Gratton \& Sneden~(1994) and Fulbright~(2000) agree well with those numbers 
at similar metallicities. A first look at these higher metallicity data 
reveals a puzzle. The average trend of [Zr/Fe] and [Y/Fe] versus [Fe/H] 
(not enough [Sr/Fe] data are available at high metallicity) are flat within 
the observational errors. However, at [Fe/H] $\sim -$0.6, the ratio [Zr/Fe] 
seems to increase slightly with decreasing [Fe/H], while [Y/Fe] seems to 
decrease slightly with decreasing [Fe/H].  
This might suggest differences in the synthesis of these two neighboring 
elements, or perhaps might be related to observational uncertainties. We will
consider this point later.

We also note the abundance scatter in the Sr, Y and Zr data. 
Specifically, starting at [Fe/H]~$\lesssim$ $-$1.0 with the 
Burris et al.~(2000) data, and becoming even more obvious at halo 
metallicities ([Fe/H] $\sim -$2), the Sr-Y-Zr abundances are characterized 
by a large [X$_i$/Fe] dispersion of $\gtrsim$ 2~dex 
(e.g., McWilliam et al. 1995; Gratton \& Sneden 1994; McWilliam 1998). 
Ultra-metal-poor stars with very large scatter of $n$-capture element abundances 
are confirmed by recent detailed studies of individual stars 
(e.g. CS~22892-052, Sneden et al.~2000a; HD~115444 and HD~122563, 
Westin et al.~2000; CS~31082-001, Hill et al.~2002). 
This large abundance scatter at the lowest metallicities, declining with
increasing [Fe/H], is usually interpreted as an early unmixed Galaxy,  
with an inhomogeneous composition of the gas in the halo 
(Gilroy et al. 1988; Burris et al. 2000).
This likely occurred because the timescale to homogenize 
the Galaxy is longer than the early evolution of the halo progenitor 
massive stars (see e.g., Ikuta \& Arimoto~1999; Raiteri et al.~1999; 
Argast et al.~2000, 2002; Travaglio et al.~2001b).  

In Fig.~6 and Fig.~7 we have plotted for our collected sample the ratios [Zr/Y], 
[Sr/Y] and [Sr/Zr] versus [Fe/H], and [Sr,Y,Zr/Ba] versus [Fe/H], respectively. 
From Fig.~4 it appears that Sr at low metallicities shows a much larger scatter 
with respect to Y and Zr. A  theoretical explanation for this effect 
is difficult. 
Nevertheless,  we notice that at [Fe/H] $< -$3, 
where the largest scatter for [Sr/Fe] is observed, there are no data available for 
[Zr/Fe] (due to observational limits) and only few data exists for [Y/Fe] (see also 
discussion in Travaglio et al.~2001b). In stars for which at least two elements among 
Sr, Y, Zr have been observed (Fig.~6) the relative dispersion is smaller.
Only the [Sr/Y] ratio (Fig.~6, middle panel) apparently shows a larger dispersion 
at [Fe/H] $\sim$ $-$3, in particular due to two major outliers, 
HD~200654 ([Fe/H] = $-$2.82, [Sr/Y] = $-$1.02, McWilliam et al.~1995,~1998) and
CS22877-011 ([Fe/H] = $-$2.92, [Sr/Y] = $-$1.55, McWilliam et al.~1995,~1998).
A relatively high [Y/Fe] in both stars seems to be the cause. 
This is also evident in Fig.~7 (middle panel) in the plot [Y/Ba] versus
[Fe/H]. From the theoretical point of view, the large scatter of Sr in 
very metal-poor stars may be naturally related to the different explosive 
properties of stars of different mass and to the consequent different 
pollution of the local interstellar medium. We recall that the $n$-component 
operating at very low metallicities is mainly due to explosive nucleosynthesis. 
Similar characteristics should also affect the $n$-component of Y and Zr. To 
determine  the possibility of an intrinsic scatter in the relative ratios Sr/Y/Zr, 
will  require further theoretical study and more observational data.

In Fig.~6 the representative points of the two selected
very $r$-process-rich stars do not show any appreciable difference with respect
to other halo stars in the sample. On the contrary, in Fig.~7 the
representative points of the two stars are situated at the lower end
of the abundance ratio spread, indicating the extremely high
$r$-process Ba content.
As will be discussed in \S~5, the interpretation of these ratios contains key 
information about the stellar origin of these elements.

Some of the stars shown in Figs.~4, 5, 6 and 7 appear to deviate from the 
mean trends enough to warrant further individual attention.
Some of these stars are like the F-star HR 107 (Edvardsson et al.~1993), 
which probably is a so-called barium dwarf that has been polluted by mass 
transfer from  a former AGB companion. 
In this case HR 107 should be taken out of our sample of field stars. 
Another special star is HD~14095 ([Fe/H] $=$ $-$0.74, Fulbright~2000); it 
has a particularly high Zr overabundance, [Zr/Fe]~= +0.58, but a solar
Y abundance, [Y/Fe]~= +0.04.  
But Fulbright (private communication) emphasizes that the Y and Zr abundances
for this star are based on very weak lines and should be viewed with caution.
The low metallicity giant HD~110184 ([Fe/H] $=$ $-$2.56, Burris et al.~2000), 
also shows high [Y/Fe] and [Zr/Fe] abundances, leading to a suspicion of AGB 
contamination. 
In fact the [Ba/Eu] value given by Burris et al.~(2000) is $+$0.31, 
much higher than the typical value for this ratio at such metallicity,
ordinarily explained by the $r$-process. 
The same caution holds for HD~105546 ([Fe/H] = $-$1.27, Burris et al.~2000), 
BD~+541323 ([Fe/H] = $-$1.65, Burris et al.~2000), and for BD~+173248 
([Fe/H] = $-$2.02, Burris et al.~2000). 
These stars have high [Sr/Fe] (+0.45, +0.57, 
+0.55, respectively), but also high [Ba/Eu] ratios (+0.10, $-$0.10, +0.01, 
respectively). We also put a warning on BD~$-$12582 and BD~+42466 (with 
[Fe/H] = $-$2.25 and [Fe/H] = $-$2.00, respectively, Burris et al.~2000). They show
a very high [Ba/Fe] ratio(+1.50 and +1.60, respectively), but no other measurements
of heavy elements are available (such a high [Ba/Fe] abundance suggests AGB contamination).
Finally, we note  two dSph stars, Fornax-21 ([Fe/H] =
$-$0.67, Shetrone et al.~2003), and Ursa Minor-K ([Fe/H] = $-$2.17, Shetrone et al.~2001).
In both cases their high [Ba/Eu] ratios ([Ba/Eu] = +0.32 and +0.33, respectively)
indicate an AGB origin.

\section{Galactic evolution of Sr-Y-Zr: results}

In this Section we first discuss the results of the Galactic evolution 
of Sr, Y, Zr, obtained using the $s$-process yields of AGB stars 
with the unweighted average over the large spread in the $^{13}$C-pocket 
efficiencies assumed (see Fig.~3 and the discussion in \S~2). An analysis of the
properties of the weak $s$-component is then afforded, as it occurs in massive
stars in a delicate balance between various nucleosynthetic sources. 
The weak $s$-component contributes a small fraction to solar Sr, 
and marginally to solar Y and Zr. No weak $s$-contribution is  expected in halo
stars, because of the strong decrease in  its efficiency with decreasing
metallicity by the effect of the secondary-like nature of the
major neutron source in massive stars, $^{22}$Ne($\alpha$,~n)$^{25}$Mg, 
and to the strong neutron poison effect of primary isotopes, like $^{16}$O, with
large abundances. We then deduce
the $r$-process isotopic fractions at the epoch of the solar system formation, 
which enables us to follow separately how the $s$- and $r$-process evolve
in the Galaxy. From the analysis of the chemical composition of peculiar very
metal-poor and very $r$-process-rich stars we quantify the primary $r$-process 
contribution to  Sr, Y and Zr that accompanies the production of the $r$-process
characterizing the heavy elements beyond Ba. This small contribution is
expected to decrease at very low metallicity, following the generally observed 
abundance decrease of the heavy neutron capture elements. The residual fraction of 
solar Sr, Y and Zr is of primary nature and is likely produced by all massive
stars; it is not strictly related to the classical $r$-process nor to the weak
$s$-process.

\subsection{$s$-Process contribution by AGB stars at solar 
system formation}

As shown in previous works on this topic (e.g., Gallino et al.~1998; 
Busso et al.~1999; Travaglio et al.~1999; Travaglio et al.~2001a), the 
main $s$-process component is clearly not the result of a `unique' astrophysical 
process.  Instead, it results from 
the integrated chemical evolution of the Galaxy,  
mixing into  the interstellar medium the output of many different generations of 
AGB stars, whose yields change with the initial metallicity, 
stellar mass, $^{13}$C-pocket efficiency and other physical properties.
This changes the traditional interpretation of the various $s$-process 
components. For example, Travaglio et al.~(2001a) showed, in the context 
of the same GCE model adopted here, that at low metallicities Pb (in particular 
$^{208}$Pb) becomes the dominant product of low-mass AGB nucleosynthesis, 
offering a natural explanation for the strong $s$-component. 

The solar abundances of Sr, Y and Zr from Anders \& Grevesse~(1989) 
are listed  in Table~1 (column 2) together with 
their uncertainties (1$\sigma$, column 3). 
The resulting Galactic $s$-fractions from AGB stars at the epoch of the solar 
system formation are listed in column 5. They were obtained taking into account the 
sum of LMS and IMS yields. As one can see from this Table, the contributions 
from IMS alone (column 4) are, in some cases, not negligible.
In particular $^{96}$Zr, an isotope originally considered of $r$-process origin
(see e.g., Cameron 1973; K\"appeler et al.~1989), receives a substantial $s$-process
contribution through the neutron capture channel at $^{95}$Zr if 
$n_n$ $\gtrsim$ 3 $\times$ 10$^{8}$ cm$^{-3}$. For a discussion of the branching 
effect at $^{95}$Zr, see K\"appeler et al.~(1990).
While IMS do not dominate the present population of AGB stars, 
they are nevertheless effective in contributing $\sim$10\% of solar Sr, Y, Zr,
with a much smaller contribution to heavier elements up to Xe, complementing the 
nucleosynthesis of LMS.

In Table~1 our corresponding $s$-process expectations for Nb and Mo are also added. 
The isotope $^{93}$Nb is bypassed during the $s$-fluence,
but its $s$-process contribution results totally from the radiogenic decay
of $^{93}$Zr. Consequently, the two elements Zr and Nb share the same 
origin -- the $s$-process contributes approximately the same amount to the 
solar abundances of each element.
Finally, Molybdenum also might be included in the Sr-Y-Zr-Nb $s$-peak, because of its 
substantial $s$-process fraction (38\% of solar Mo). For this reason, we have 
tentatively included Mo in Table~1.
In the list of isotopes reported in Table~1 we excluded the p-only isotopes $^{84}$Sr and 
$^{92,94}$Mo, which are bypassed by the $s$-process. Note that for Mo this corresponds to 
a contribution of 14\% to the solar value. We also excluded the $r$-only $^{100}$Mo. 
We note that $^{86,87}$Sr and $^{96}$Mo are $s$-only isotopes and from our total 
$s$-process (main+weak) predictions for the solar composition (Table~1, column 7), 
we obtain total contributions of 
76\%, 70\%, and 80\%, respectively. Therefore, an additional 
contribution from 
a {\it slow} neutron-capture process is needed to reproduce their solar composition.

The uncertainties in the $s$-fractions depend on the prescriptions 
for AGB yields and the GCE model, as well as on the uncertainty of 
neutron capture cross sections and solar abundances. 
As for the experimental neutron capture cross 
sections, those for the  Sr isotopes and for  $^{89}$Y are fairly well known, at a 
5\% level 
or less (Bao et al.~2000; Koehler et al.~2000). A similar precision has been
achieved for $^{94,96}$Zr by Toukan \& K\"appeler~(1990), whereas for 
$^{90,91,92,93}$Zr a higher uncertainty of about 10\% is
reported in the recent compilation of cross sections by Bao et
al. (2000). These latter cross sections are 
based upon the older measurements of Boldeman et al.~(1976).
A further uncertainty affects the $^{96}$Zr $s$-process contribution,
which is fed by the neutron channel at $^{95}$Zr, whose neutron capture
cross section is based on theoretical estimates only.
A reliable determination  of these cross sections with improved 
measurements is highly 
desirable. In column 6 we report the contributions to these isotopes 
from the weak $s$-component, according to the analysis of 
Raiteri et al.~(1993; see discussion in \S~5.2). 
Finally, in column 7 we report the total $s$-process contributions
from AGBs and from the weak $s$-component in massive stars.

In the second column of Table~2 we report the updated predictions from the
classical analysis (Arlandini et al.~1999: `classical'').
In column 3 we report the $s$-process predictions for the best fit to the 
main $s$-component (indicated as `stellar model' in Arlandini et al.~1999).
Those authors obtained this result with AGB models with masses 
from 1.5 to 3 \msb, a metallicity of [Fe/H] $=$ $-$0.3 and with  
the ST choice for the
$^{13}$C-pocket. A  comparison with the results shown in Table~2 makes clear 
that, even using the same updated neutron capture network, the classical
analysis does not allow any $r$-process residuals for Y (at odds with
spectroscopic observations of low metallicity stars). For the light $s$-elements, 
both the 'classical analysis' and the 'stellar model' by Arlandini et al.~(1999) give 
different prescriptions with respect to the ones obtained by 
integrating AGB $s$-yields
over metallicity in the framework of the GCE model.

In Table~3 we report our GCE predictions for all elements from Co up to
Mo, taking into account the weak $s$-process from Raiteri et
al.~(1992) (column 2), our LMS and IMS predictions separately (columns 3 and 4), 
and the total predictions from AGB stars  (column 5). Finally, our predictions for 
the total $s$-process (IMS + LMS + weak-$s$ component) fractions 
of solar abundances are reported in column 6.
From Table~3 we see  that solar Kr and Rb have an $s$-fraction of $\sim$50\%. 
Those two elements belong to the first $s$-peak at N = 50.
Also solar gallium and germanium have a contribution from 
the $s$-process of $\sim$50\%,
while for  selenium the $s$-process is responsible for 
$\sim$40\%. The major $s$-component that contributes to the solar abundance 
of these three elements comes from the metallicity-dependent
weak $s$-process. We also notice that among the two easily
observable and nearby elements copper and nickel, only copper
is affected by the $s$-process at  $\sim$30\%. This is mostly due to 
the weak
$s$-process in massive stars. Nickel is almost unaffected by the
$s$-process (see Raiteri et al.~1992; Matteucci et al.~1993; 
Mishenina et al.~2002; Simmerer et al.~2003).

\begin{table}
\begin{center} TABLE 1\\
{\sc $s$-process fractional contributions at {\it t} = {\it t}$_{\odot}$}\\
{\sc with respect to solar system abundances}\\
\vspace{1.0em}
\begin{tabular}{r|cc|ccc|c}
\hline\hline
  & Solar $^{(\rm a)}$ & & GCE$^{(\rm b)}$ &  GCE$^{(\rm b)}$ 
& Weak-$s$$^{(\rm c)}$ & TOT-$s$$^{(\rm d)}$ \\
  &  (atom) &  ($\sigma$)   & (IMS) & (LMS+IMS) &  &  \\
  &  (\%)  & (\%) & (\%) & (\%) & (\%) & (\%) \\
\hline

$^{86}$Sr & 9.86  &  & 8  & 52 & 24 & 76 \\
$^{87}$Sr & 7.00  &  & 5  & 54 & 16 & 70 \\
$^{88}$Sr & 82.58 &  & 10  & 75 & 7 & 82 \\
 {\bf Sr} &    & 8.1 &  {\bf 9} & {\bf 71} & {\bf 9} & {\bf 80} \\

\hline

$^{89}$Y  & 100 &  & 7  & 69 & 5 & 74 \\
 {\bf Y} &    & 6.0  & {\bf 7} & {\bf 69} & {\bf 5} & {\bf 74}  \\

\hline

$^{90}$Zr & 51.45 &  &  6 & 53 & 2 & 55 \\
$^{91}$Zr & 11.22 &  & 18 & 80 & 3 & 83 \\
$^{92}$Zr & 17.15 &  & 15 & 76 & 3 & 79 \\
$^{94}$Zr & 17.38 &  &  9 & 79 & 2 & 81 \\
$^{96}$Zr & 2.80  &  & 40 & 82 & 0 & 82 \\
 {\bf Zr} &    & 6.4 & {\bf 10} & {\bf 65} & {\bf 2} & {\bf 67}  \\

\hline

$^{93}$Nb  & 100 &  & 12  & 67 & 2 & 69 \\
 {\bf Nb} &    & 1.4 & {\bf 12} & {\bf 67} & {\bf 2} & {\bf 69} \\

\hline
$^{95}$Mo  & 15.92 &  & 4  & 39 & 1 & 40 \\
$^{96}$Mo  & 16.68 &  & 8  & 78 & 2 & 80 \\
$^{97}$Mo  & 9.55 &  & 6  & 46 & 1 & 47 \\
$^{98}$Mo  & 24.13 &  & 6  & 59 & 1 & 60 \\
 {\bf Mo} &    & 5.5 & {\bf 4} & {\bf 38} & {\bf 1} & {\bf 39} \\

\hline
\hline
\end{tabular}
\end{center}

\hspace{9em}$^{(\rm a)}$ -- Anders \& Grevesse~(1989)
\vspace{0.1em}

\hspace{9em}$^{(\rm b)}$ -- This work
\vspace{0.1em}

\hspace{9em}$^{(\rm c)}$ -- Raiteri et al.~(1993)
\vspace{0.1em}

\hspace{9em}$^{(\rm d)}$ -- Total from $s$-process: main-$s$ + weak-$s$
\vspace{0.1em}

\end{table}

\begin{table}
\begin{center} TABLE 2\\
{\sc $s$-process fractional contributions at {\it t} = {\it t}$_{\odot}$}\\
{\sc with respect to solar system abundances}\\
\vspace{1.0em}
\begin{tabular}{rccc}
\hline\hline
  & Main-$s$$^{(\rm a)}$ & Main-$s$$^{(\rm b)}$ & GCE \\
  &  (\%)  & (\%) & (\%) \\
\hline

$^{86}$Sr & 68 & 47 & 52\\
$^{87}$Sr & 74 & 50 & 53\\
$^{88}$Sr & 94 & 92 & 75\\
 {\bf Sr} & {\bf 90} & {\bf 85} & {\bf 71}\\

\hline

$^{89}$Y  & 106 & 92 & 68\\
 {\bf Y}  & {\bf 106} & {\bf 92} & {\bf 69}\\

\hline

$^{90}$Zr & 68 & 72 & 56\\
$^{91}$Zr & 100 & 96 & 88\\
$^{92}$Zr & 108 & 93 & 82\\
$^{94}$Zr & 116 & 108 & 84\\
$^{96}$Zr & 51 & 55 & 101 \\
 {\bf Zr} & {\bf 82} & {\bf 83} & {\bf 65}\\

\hline

$^{93}$Nb & 100 & 85 & 67\\
 {\bf Nb} & {\bf 100} & {\bf 85} & {\bf 67}\\

\hline
$^{95}$Mo & 55 & 55 & 39\\
$^{96}$Mo & 116 & 106 & 78\\
$^{97}$Mo & 68 & 59 & 46\\
$^{98}$Mo & 90 & 76 & 59\\
 {\bf Mo} & {\bf 54} & {\bf 50} & {\bf 38}\\

\hline
\hline
\end{tabular}
\end{center}

\hspace{9em}$^{(\rm a)}$ -- Arlandini et al.~(1999), ``classical analysis''
\vspace{0.1em}

\hspace{9em}$^{(\rm b)}$ -- Arlandini et al.~(1999), ``stellar model''
\vspace{0.1em}

\hspace{9em}$^{(\rm c)}$ -- GCE $s$-process fraction LMS+IMS
\vspace{0.1em}

\end{table}

\begin{table}
\begin{center} TABLE 3\\
{\sc $s$-process contribution at {\it t} = {\it t}$_{\odot}$}\\
\vspace{1.0em}
\begin{tabular}{r|c|ccc|c}
\hline\hline
  & Massive stars$^{(\rm a)}$     &    &  AGB stars$^{(\rm b)}$  &  &   TOT-$s$  \\
  & Weak-$s$& LMS &  IMS  &  TOT AGB  &  
Weak-$s$+Main-$s$ \\
  &  (\%)  & (\%) &  (\%) &  (\%) &  (\%) \\
\hline

{\bf Co} &  6  &  1  &  2  &  3  &  9 \\
{\bf Ni} &  1  &  0  &  0  &  0  &  1 \\
{\bf Cu} & 22  &  2  &  3  &  5  & 27 \\
{\bf Zn} &  8  &  2  &  1  &  3  & 11 \\
{\bf Ga} & 44  &  7  &  4  & 11  & 55 \\
{\bf Ge} & 43  &  8  &  4  & 12  & 55 \\
{\bf As} & 17  &  5  &  3  &  8  & 25 \\
{\bf Se} & 25  &  9  &  5  & 14  & 39 \\
{\bf Br} & 11  &  9  &  6  & 15  & 26 \\
{\bf Kr} & 19  & 17  & 12  & 29  & 48 \\
{\bf Rb} & 14  & 18  & 21  & 39  & 53 \\
{\bf Sr} &  9  & 62  &  9  & 71  & 80 \\
{\bf Y } &  5  & 62  &  7  & 69  & 74 \\
{\bf Zr} &  2  & 55  & 10  & 65  & 67 \\
{\bf Nb} &  2  & 55  & 12  & 67  & 69 \\
{\bf Mo} &  1  & 34  &  4  & 38  & 39 \\

\hline
\hline
\end{tabular}
\end{center}

\hspace{9em}$^{(\rm a)}$ -- Raiteri et al.~(1992)
\vspace{0.1em}

\hspace{9em}$^{(\rm b)}$ -- This work
\vspace{0.1em}

\end{table}

In Fig.~8 we show ({\it full dots}) the Galactic chemical 
contribution at the epoch of the solar system formation of elements 
in the atomic number range Z = 6 to Z = 82, taking into account AGBs of low and
intermediate mass. For the light elements below Fe,
there is a small $s$-process AGB contribution
to  P (2.1\%) and Sc (1.6\%). Also AGBs make large contributions 
to the isotopes $^{12}$C (29\% solar) and $^{22}$Ne
(44\% solar) leading to  total element fractions of 
29\% for  solar carbon and 3.5\% for  solar neon (Arnone et al.~2003).
As to nitrogen, we only considered the yields during the advanced TP-AGB phase when 
the star suffers thermal pulses and third dredge up episodes. Here, all the $^{14}$N 
nuclei produced in the H-burning shell by the full operation of the HCNO cycle are 
subsequently converted by double $\alpha$ capture into $^{22}$Ne. However, a major 
contribution to solar N comes from LMS in the Red Giant phase due to the 
first dredge up. During this phase, material of the inner 
radiative zones is mixed with the surface by the extension of the convective envelope, 
and proton captures convert about 1/3 of the initial $^{12}$C into $^{14}$N. 
Further substantial contribution from the nitrogen content in the Galaxy derives from 
the operation of the so-called `cool bottom process' in low-mass AGBs, and of the 
`hot bottom process' in intermediate mass stars (see discussion in Busso et al.~1999 and 
references therein). Both contributions have not been considered here.

For elements beyond Fe and up to Zn, AGBs make minor $s$-process contributions
to Co (2.7\%), Cu (5.2\%) and Zn (2.6\%).
For comparison we also show, for elements from Sr up to Bi, 
the $s$-process predictions for the best-fit to the main $s$-component 
from Arlandini et al. (1999): from the `stellar model' ({\it open squares}; also Table~2, 
this work) and the updated predictions by the classical analysis (Arlandini
et al.~1999: `classical analysis'; also Table~2, this work).
It is clear from this figure  that in the Ba--Eu region, and up to Tl, the GCE
model agrees quite well with the Arlandini et al.~(1999) `stellar model'. 
This is mainly due to the fact that the metallicity region around 
[Fe/H] = $-$0.3, corresponding to the stellar models adopted by 
Arlandini et al.~(1999), is also the metallicity where the higher AGB 
production of elements between Ba--Eu and up to Tl occurs 
(Travaglio et al.~1999).
The discrepancy between the predictions for Pb and Bi (see Travaglio 
et al.~2001a for a detailed discussion) from the `classical analysis' 
and from the `stellar model' is a clear indication that neither a 
unique AGB nor the classical analysis is able to explain the main 
$s$-component in the solar system. As a matter of fact, the main $s$-component 
is the outcome of different generations of AGB stars prior to the solar 
system formation. 

\subsection{The weak $s$-process and its contribution to solar Sr, Y, Zr}

Neutron capture processes in massive stars at different metallicities 
play a role in the Galactic production of Sr, Y, and Zr. 
We intend to discuss in this Section some relevant points of this problem.

As we already noted  in \S~1, the reaction $^{22}$Ne($\alpha$,~n)$^{25}$Mg 
represents the major neutron source for the weak $s$-process in massive stars. 
It takes place partly in the final phases of core He-burning (near He exhaustion)
when the central temperature rises up to 3.5 $\times$ 10$^8$ K, and partly in the 
subsequent convective C-burning shell at a much higher temperature 
(around 1 $\times$ 10$^9$ K). There, a copious
release of $\alpha$ particles comes out from the major reaction
channel $^{12}$C + $^{12}$C $\rightarrow$ $^{20}$Ne + $\alpha$.
Consequently,  this  $s$-process production has a complex dependence on 
the initial mass. Indeed, during core He burning, $^{22}$Ne
is less consumed in the less massive stars, and thus a major
fraction is available for the subsequent convective shell C-burning phase.
In the more massive stars, almost all $^{22}$Ne  has been consumed
already by  core He exhaustion, and a saturation in the neutron
exposure is reached (Prantzos et al. 1990).
During convective core He burning, the neutron density barely reaches 1
$\times$ 10$^6$ cm$^{-3}$. In contrast  at the beginning of 
convective shell
C-burning the neutron density shows a sharp exponential decline
from an initial very high value, of $\sim$1 $\times$ 10$^{11}$ cm$^{-3}$
(Arnett \& Truran 1969; Raiteri et al. 1991b, 1993), due to the early release
of $\alpha$ particles.

Previous analyses of the weak $s$-component (e.g., Couch, Schmiedekamp, 
\& Arnett 1974;  Lamb et al. 1977; Prantzos et al. 1990; Raiteri et al. 1991a,
1992, 1993) provided a decreasing $s$-process contribution from  massive stars 
with increasing atomic number, on  the 
order of 70\% for the $s$-only $^{70}$Ge and of 30\% for the $s$-only $^{76}$Se.
As for Kr, the extreme temperature dependence of the 
mean-life of $^{79}$Se favors the production of the $s$-only  $^{80}$Kr 
(this isotope is mostly produced, by 80\%, in massive stars). 
Also the $s$-only $^{82}$Kr receives an important contribution 
(50\%) from  
the weak $s$-component. 
Concerning the Sr isotopes, the branching effect
at $^{85}$Kr favors the $s$-only $^{86}$Sr and $^{87}$Sr, whose solar abundances
are produced  by about 20\% from the weak $s$-process.
The low overall neutron exposure almost stops the $s$ fluence at the neutron 
magic $^{88}$Sr. This isotope receives only a $\sim$5\% contribution 
from the weak $s$-process. From $^{89}$Y on, the weak $s$-process 
contribution is marginal. 

It should be clear from the previous considerations that the classical analysis 
of the weak $s$-process, which typically requires a constant temperature and 
constant neutron density and an unknown distribution of neutron exposures, 
is not suitable at all in approximating  the weak $s$-process occurring in massive 
stars.

In massive stars a general note of caution needs to 
 be addressed to the sensitivity 
of the predicted abundances of nuclei at neutron magic N = 50 and 
on the uncertainty 
in the cross section of several lighter isotopes for which only theoretical 
estimates are given, including  $^{62}$Ni, $^{72,73}$Ge and $^{77,78}$Se. 
Moreover, many cross sections of stable isotopes from Fe to Sr  still have  
uncertainties of at least  10\%.
Also the temperature dependence of the cross section of several isotopes shows
strong departures from the usual $1/v$ law, among them  $^{56}$Fe, 
$^{61}$Ni, $^{63}$Cu, $^{67}$Zn, $^{71}$Ga, $^{73}$Ge, $^{75}$As, and all the Kr 
isotopes. 

Finally, several primary light isotopes act as major neutron absorbers
 at 
low metallicity. Their cross sections often show large departures from the 
$1/v$ law and, therefore, need to be carefully 
evaluated. Among them are $^{12}$C, $^{16}$O, $^{20}$Ne and all Si isotopes 
(see Bao et al.~2000). 
Of particular importance for the $s$-process efficiency 
is $^{16}$O, for which Igashira et al.~(1995) measured
a neutron capture cross section that turned out to be 170$\times$
higher than previous theoretical estimates by Allen \& Macklin~(1971).
As a matter of fact, even employing the much lower value from  Allen \&
Macklin~(1971), the neutron capture on $^{16}$O in massive stars was found to
strongly depress the weak $s$-process at low metallicity (Raiteri et
al.~1992). The effect may be moderated by the partial recycling effect of 
the chain $^{16}$O(n, $\gamma$)$^{17}$O($\alpha$,~n)$^{20}$Ne (Travaglio
et al.~1996). Recently, Woosley et al.~(2003) noted  the importance 
of including  the effect of neutron poisoning from $^{16}$O
in core collapse supernovae of solar metallicity.

Besides all of the above intricacies, a major impact is played by 
the need for  a
careful knowledge of key reaction rates, like $^{22}$Ne($\alpha$,~n)$^{25}$Mg
and $^{22}$Ne($\alpha$,~$\gamma$)$^{26}$Mg (K\"appeler et al.~1994; see Rauscher
et al.~2002; Woosley et al.~2003).
One has finally to recall the critical effect  on the advanced phases of stellar
evolution and nucleosynthesis that results  from the choice for  the important,
but uncertain,   
$^{12}$C($\alpha$,~$\gamma$)$^{16}$O reaction rate
(see Rauscher et al.~2002 for a recent discussion). Other
difficulties are related to a realistic treatment of convective-radiative borders,
of the time-dependent mixing and nucleosynthesis processes, of the inclusion
of rotation and mass loss, as well as on hydrodynamic
multi-dimensional effects.

In spite of the above mentioned problems affecting neutron capture nucleosynthesis in 
massive stars and the nucleosynthetic origin of the $r$-process,
a general conclusion may be drawn. Due to the metallicity
dependence of the major neutron source, $^{22}$Ne, no contribution from the
weak $s$-process is to be expected in halo stars. Some extra contribution is 
however to be expected from primary neutron sources in massive stars, whose quantitative 
impact is difficult to determine at present.

\subsection{Primary neutron capture sources in massive stars not related
to  the classical $r$-process}

In the more advanced evolutionary stages of massive stars there are 
a few 
primary neutron sources that may be activated, among which is the possible 
contribution during carbon burning of  the sub-threshold channel
$^{12}$C + $^{12}$C $\rightarrow$ $^{23}$Mg + n (Caughlan \& Fowler~1988).
Another important source of neutrons during C-burning is the reaction
$^{26}$Mg($\alpha$,~n)$^{29}$Si, where $^{26}$Mg is partly of primary origin
from the chain $^{12}$C + $^{12}$C$ \rightarrow$ $^{23}$Na + p followed by
$^{23}$Na($\alpha$,~p)$^{26}$Mg. In the subsequent hydrostatic O-burning phase, 
which takes place at around 2 $\times$ 10$^9$ K, an intense primary neutron
production is released by the reaction channel $^{16}$O + $^{16}$O
$\rightarrow$ $^{31}$S + n. Explosive nucleosynthesis governs the 
yields of the ejecta of the Si-rich zone and of the inner region of the O-rich 
zone, where photodisintegration processes on heavy isotopes play a consistent 
role on  dynamical time scales. In the post-explosive nucleosynthesis 
yields of massive stars with solar composition (Rauscher et al.~2002), with 
an expected sharp decline for nuclides beyond the neutron magic N = 50, there 
are comparable amounts of $r$-only and $s$-only isotopes, 
$^{70}$Zn and $^{70}$Ge, $^{76}$Ge and $^{76}$Se, $^{80}$Se and $^{80}$Kr, 
$^{82}$Se and $^{82}$Kr, $^{86}$Sr and $^{86}$Kr, which is impossible to explain 
either by the  weak $s$-process, or by  a pure $r$-process mechanism. 

These first detailed results for the build-up of heavy elements with a full
reaction network in core collapse supernovae are reminiscent of the older
numerical simulations aimed at characterizing the astrophysical site for the
nucleosynthesis of the $r$-process. For example, simulations of neutron
captures occurring after the passage of a shock front at explosive He-shell
conditions ($\rho$ $\lesssim$ 10$^5$ g cm$^{-3}$, 
0.9 $\times$ 10$^9$ K $\lesssim T \lesssim 2 \times 10^9$ K) were able to 
reproduce only the $r$-process peak at A $=$ 80 (Hillebrandt, 
Takahashi, \& Kodama~1976; Hillebrandt, \& Thielemann~1977; Truran, Cowan, \& 
Cameron~1978; Cowan, Cameron, \& Truran~1985). That result was considered a failure 
in the quest for  a common astrophysical site for reproducing the whole $r$-process
distribution in the solar system, from A $\sim$ 80 up
to the transuranics. Nevertheless, the growth of spectroscopic
data now available suggests that neither  the $s$-process, nor the $r$-process in nature 
are the result of unique nucleosynthesis processes.

\subsection{The $r$-process contribution to Sr, Y, Zr as deduced from very 
metal-poor and very $r$-process-rich stars}

\begin{table}
\begin{center} TABLE 4\\
{\sc $s$-process contribution and $r$-process fraction at the solar
composition}\\
\vspace{1.0em}
\begin{tabular}{rc|ccc}
\hline\hline
  & $s$-fraction &  $r$-fraction  &  $r$-fraction   &  $n$-fraction \\
  & AGB+weak-$s$  &  ($r$-residuals) & (from CS 22892-052) & \\
  &  (\%)  & (\%) &  (\%) &  (\%)\\
\hline

{\bf Sr} & 80  &  20  &  12  &  8 \\
{\bf Y } & 74  &  26  &  8   &  18 \\
{\bf Zr} & 67  &  33  &  15  &  18 \\
{\bf Nb} & 69  &  31  &  13  &  18 \\
{\bf Mo} & 39  &  37  &  12  &  25 \\

\hline
\hline
\end{tabular}
\end{center}

\end{table}

\begin{table}
\begin{center} TABLE 5\\
{\sc $s$-process contribution and $r$-process fraction at the solar
composition for elements from Ru to Cd}\\
\vspace{1.0em}
\begin{tabular}{rc|cc}
\hline\hline
  & $s$-fraction &  $r$-fraction  &  $r$-fraction   \\
  & AGB+weak-$s$  &  ($r$-residuals) & (from CS 22892-052) \\
  &  (\%)  & (\%) &  (\%)  \\
\hline

{\bf Ru} & 24  &  69  &  50  \\
{\bf Rh} & 10  &  90  &  43  \\
{\bf Pd} & 36  &  64  &  36  \\
{\bf Ag} &  9  &  91  &  30  \\
{\bf Cd} & 38  &  62  & (41)  \\
\hline
\hline
\end{tabular}
\end{center}

\end{table}

As briefly described in \S~3.2, our estimate of $r$-process abundances
for the elements from Ba to Pb at $t=t_\odot$ has been derived 
using  the
$r$-process residuals method. If we would apply the same method to Sr-Y-Zr,
we would obtain 
an $r$-process contribution of $\sim$20--30\% to the solar 
composition (see Table~3). Nevertheless, the more complex nucleosynthetic 
origin of the Sr-Y-Zr elements with respect to the Ba-Eu elements is suggested 
by $r$-process-rich and very low metallicity stars, such as CS~22892-052.
Since this star has an $r$-process enrichment of $\sim$40$\times$ the 
solar-scaled composition (see Fig.~5 [Eu/Fe] versus [Fe/H], middle panel), 
we can ascribe the signature
of CS~22892-052 to the `pure' $r$-process (i.e., any contamination by other 
possible stellar sources is hidden by the $r$-process abundances).
A very similar trend in both Sr-Y-Zr and in the heavy elements
beyond Ba is shown by another very $r$-process-rich star of nearly the 
same metallicity, e.g., CS~31081-001. Both CS~22892-052 and CS~31081-001 are 
highlighted as bold symbols in the various Figures.
In particular, these two stars show the highest [Sr/Fe], [Y/Fe], and [Zr/Fe] 
ratios among stars of comparable metallicity (see Fig.~4).
The other special stars indicated as open circles, for which accurate 
spectroscopic abundances of many neutron-capture elements are available, 
do not show such an extreme $r$-process-rich signature.
For them, both Sr, Y and Zr, or  Ba, Eu cannot be
distinguished from the averages of other stars of comparable metallicity.   

Under the above assumption, and knowing that the Ba $r$-fraction at the epoch 
of the solar system formation is $\sim$20\% (Travaglio et al.~1999; 
Arlandini et al.~1999), from CS~22892-052 one can derive the $r$-fraction for  
Sr-Y-Zr of $\sim$10\% (and {\it not} $\sim$25--30\% as derived from the $r$-residuals 
method). Note that the same result can be obtained employing  Eu instead of Ba. 
In addition, as noted  above, we know that the $s$-process contributes 
80\%, 74\%, and 67\% to solar Sr, Y and Zr, respectively (see Table~4, second column). 

After summing up all these contributions we find that fractions of 8\%, 
18\%, and 18\% of solar Sr, Y, and Zr, respectively, are ``missed''. 
We then assume that this missing fraction is of `primary' origin and results from all 
massive stars. We note that at this time it is not possible to completely define this 
additional nucleosynthetic contribution to Sr, Y and Zr.  In  
the advanced stages of evolution in massive stars, and in particular
during explosive oxygen burning, a number of `primary' neutron sources
can be activated. The situation is complicated, however, and neutron
production might also be accompanied by 
photodisintegrations, as well as by proton and alpha captures. 
It appears  that this nucleosynthesis 
is only contributing to the production of 
the lighter $n$-capture elements (Sr-Zr) -- although  
the production of all elements from Cu to Zr could be affected -- and this is of
a `primary' nature. For ease of discussion we would label this additional
nucleosynthesis as a lighter element primary process (LEPP). 
We emphasize  further that detailed (full network) 
supernova model calculations,
for stars of low metallicity (as opposed to solar-metallicity models)
are not yet available but will be required to better understand this
nucleosynthesis.  Additional observational data, particularly for 
Sr-Y-Zr in low metallicity stars, will also help to constrain 
theoretical models and better define the nature of lighter $n$-capture element
synthesis in low-metallicity stars.
The fractional contributions to Sr-Y-Zr from this LEPP 
are reported in the last column of Table~4. In Table~4 (column 3) we also 
report for comparison purposes 
the $r$-fraction obtained with the $r$-residuals method. 
In the case of Nb, since its abundance results
 from the radiogenic decay of $^{93}$Zr, 
we have adopted the same contribution estimated for Zr. Therefore, knowing 
its total $s$-process contribution to solar Nb, we have deduced the $r$-fraction 
without relying on the still uncertain observed value of [Nb/Fe].   
We also warn the reader on the Mo $r$-fraction derived from the same star, due 
to  difficulties in detecting 
 Mo in low-metallicity stars (see discussion in Sneden et al.~2003a, 
where the error bar for the Mo abundance in CS~22892-052 has been estimated to 
be $\sim$0.2 dex).

In Table~5 we derive from CS~22892-052 the $r$-fraction of elements from Ru to Cd 
strictly based on the production of the heavy $r$-elements beyond Ba.  
In the case of Ru a $p$-process contribution (to $^{96}$Ru and $^{98}$Ru) has been 
taken into account, affecting the solar Ru by $\sim$7\%. 
These elements are not the subject of this work, and therefore we 
will not enter into a 
detailed discussion here. Since Cd in the CS~22892-052 is an upper limit 
we report in 
parenthesis the $r$-fraction derived for that element in this star. 

The discrepancy in the $r$-fraction 
of Sr-Y-Zr between the $r$-residuals method and  the CS22892-052 abundances 
becomes even larger for elements from Ru to Cd -- the weak $s$-process does not
contribute to elements from Ru to Cd. As noted in the Introduction, this 
discrepancy suggests an even more complex multisource nucleosynthetic origin 
for elements like Ru, Rh, Pd, Ag, and Cd.

In Fig.~9 the $n$-capture abundance 
pattern observed in the ultra-metal-poor star CS~22892-052 is compared 
with the $r$-process abundance curve obtained by computing the
Galactic chemical evolution of $s$-process nucleosynthesis from AGB
stars, and using the $r$-process residuals method to infer the
$r$-process fractions. Exceptions are the $r$-process contributions to 
Sr, Y, Zr, Nb, and Mo, for which we derived the $r$-process fraction 
from CS~22892-052. The values plotted in Fig.~9 correspond to 
the ones reported in Table~4, column 4. Even when 
we derived the $r$-fraction 
for Sr, Y, Zr, Nb and Mo 
from the CS~22892-052, small differences between 
our theoretical prediction and the observational data are still visible.
This is due to observational uncertainties affecting the data
(e.g., deriving the $r$-fraction using Eu instead of Ba will 
results in  differences 
up to 5\% for the elements from Sr to Mo).

Spectroscopic data of the elemental abundances in CS~22892-052 
(Sneden et al.~2003a) plotted in Fig.~9 are the 
result of an average between ground-based  observations and HST observations (with 
the exception of Pb). For Pb both ground-based data as well as HST data have 
been plotted. For Au no ground-based 
data are available, therefore only HST data have been plotted.
Except for Pb, all HST data (Ge, Y, Os, Pt, Au, Pb) seem to be consistent 
with the ground-based data. Concerning Pb, the uncertainties are still 
significant even in the ground-based data, as Sneden et al.~(2003a) commented 
in their work
" ... we are not confident of the suggested detections of two 
\ion{Pb}{1} lines in the ground-based spectra."
We note that for Nb and Mo the observational error bars plotted are `adopted 
error bars' (see Sneden et al.~2003a for discussion).
We also added error bars to our theoretical GCE predictions. They are based, 
for each isotope, on uncertainties in the solar system elemental abundances 
(Anders \& Grevesse~1989), and on the uncertainties in neutron capture cross 
sections (Bao et al.~2000).
These error bars are  important for elements
belonging to the three $s$-peaks, and in particular for Pb.
In the case of La, we have adopted  a conservative 2$\sigma$ uncertainty,
since discrepant experimental determinations of the neutron cross section 
are reported in the compilation of Bao et al.~(2000). 

\section{Reconstruction of the Galactic evolution of Sr, Y, Zr, 
by diverse neutron capture mechanisms}
 
In Fig.~10, we show the Galactic evolutionary trends predicted by our 
model for [Sr/Fe], [Y/Fe], [Zr/Fe] versus [Fe/H].
These predictions take into account the $s$-process, the
$r$-process and the primary process (or LEPP), 
in the halo, thick-disk and thin-disk.
In Fig.~10 also the GCE model predictions for the $s$-process contribution by
AGB stars alone are reported.
For Galactic disk metallicities, in the predicted total ($s$ + $r$ + $primary$) 
average trend one can distinguish the effect of the late AGB contribution, and also 
a tiny difference in the relative behavior of Sr, Y and Zr at [Fe/H] $>$ $-$0.3.
It is tempting to interpret in this manner the different trends 
observed in [Zr/Fe] and [Y/Fe]  
in disk metallicity stars as described  in the Introduction.
However, a more detailed analysis would be needed when comparing results 
of different 
authors in order to distinguish intrinsic abundance variations 
from observational uncertainties. 

In Fig.~11 we show the predicted trends for [Sr/Y], [Sr/Zr] and [Y/Zr]
versus [Fe/H], and in Fig.~12 the predicted trends for [Sr/Ba], [Y/Ba],
[Zr/Ba] versus [Fe/H]. In particular, we focus on the ratio of Sr, Y, Zr 
over Ba for the following reason. Ba, as discussed previously, at low metallicity 
is mainly produced by $r$-process nucleosynthesis. {\it If} Ba and Sr-Y-Zr would 
derive from the same stellar source at such low metallicities, we would expect 
to see a fairly flat ratio vs. [Fe/H] (within observational error bars), as in the 
case of [Ba/Eu] (see Travaglio et al.~1999 for further discussion). 
This does not seem to be the case for Sr-Y-Zr, suggesting a different 
stellar origin for Sr-Y-Zr and Ba.

At very low metallicity the increasing [Sr,Y,Zr/Ba] ratio with decreasing [Fe/H]
is correlated with the delayed production of Ba with respect to Sr-Y-Zr. This 
assumes  that at low metallicity Ba is produced in type II Supernovae 
in the mass range 8 $-$ 10 $M_\odot$, while Sr-Y-Zr derive their $n$-component 
from all massive stars. The $r$-fraction alone is shown in Fig.~12 as a flat line 
in the three panels, which is consistent with the values for 
 the two extremely $r$-process-rich stars
CS~22892-052 and CS~31082-001.

There have been recent observations of $n$-capture elements in nearby 
dwarf spheroidal galaxies (Shetrone et al.~2001,~2003). 
Abundance comparisons in these galaxies, as a function of metallicity, 
demonstrate similar patterns to those  observed in the Galaxy. 
From the Shetrone et al. data for Ba and Eu (our Fig.~5) we note that 
the abundance scatter present in [Ba/Fe] and [Eu/Fe] for [Fe/H] $<$ $-$1.5 
disappears at increasing metallicities, in a manner similar to the 
Galactic sample. Such an agreement among these various galaxies suggests a 
common synthesis history for these elements as a function of iron production.
Nevertheless, the [Y/Fe] in dwarf galaxies stars with $-$2 $<$ [Fe/H] $<$ $-$1 seems 
on average lower than what is observed in the Galaxy (Fig.~4 and Fig.~10). 
The same is true for the ratio [Y/Ba] (Fig.~7 and Fig.~12). 
Of course the stellar sample in dwarf galaxies is still too low to draw any final
conclusions. Moreover, no data for Sr and Zr are available, and different 
star formation
histories for each dwarf galaxy should be carefully taken into account (Tolstoy et 
al.~2003). In spite of these intricacies, one possible explanation for the observed 
lower [Y/Fe] and [Y/Ba] ratios in dSph galaxies is that less ejecta from 
massive stars are retained in the local interstellar medium,  consequently reducing 
the contribution to Y from a primary process (see discussion in this 
paper). Summing up, an increase in  the small number statistics of stars observed in 
dSph galaxies, as well as additional 
data for Sr and Zr in dSph stars, will allow us to better 
understand the nucleosynthesis history in these galaxies.

\section{Conclusions}
In this paper we have calculated the evolution of the light $n$-capture 
elements, Sr, Y, and Zr. The input stellar yields for these nuclei have been 
separated into their $s$-, $r$-, and primary-process components. The $s$-yields 
are the result of post-process nucleosynthesis calculations based on full 
evolutionary AGB models computed with the FRANEC code. Spectroscopic observations 
of very low-metallicity stars in the Galaxy,  as well as the first observations of 
single stars in dwarf spheroidal galaxies, suggest that an extra source (of primary 
nature) is needed to synthesize Sr, Y, and Zr, to enrich the early interstellar 
medium, and to reproduce the solar composition of $s$-only isotopes 
like $^{86,87}$Sr, 
$^{96}$Mo. 
We therefore think that neutrons should give the major imprint to this primary
process, that has to be considered different from the `classical $s$' and the `classical $r$' 
processes. The results of the Galactic evolution model confirm these 
observational indications. 

We compared our theoretical predictions with the abundance pattern observed 
in the very $r$-process-rich  CS~22892-052 (Sneden et al.~2003a). This star 
is known to show a {\it pure} $r$-process signature (it shows a $r$-process 
enhancement of about 40$\times$ the solar value, much larger than any abundance 
observed in normal halo stars). We extracted from this star the $r$-fraction
of Sr, Y, and Zr ($\sim$10\% of the solar value). In the light of our 
nucleosynthesis calculations in AGB stars at different metallicities, 
integrated over the GCE model briefly described in this paper, we conclude 
that the $s$-process from AGB stars contributes to the solar abundances 
of Sr, Y and Zr 
by  71\%, 69\% and 65\%, respectively. To the solar Sr abundance, we also added
a small contribution ($\sim$10\%) from the `secondary' weak $s$-component from 
massive stars.

As a consequence of the above results, we conclude that a primary component from 
massive stars is needed to explain 8\% of the solar abundance of Sr, and 18\% of 
solar Y and Zr. Although this contribution to  the solar composition is small, 
especially in terms of  the overall uncertainties, it nevertheless appears to be 
necessary to produce the observed enrichment of these elements in the very  
low-metallicity stars. This process is of {\it primary} nature, unrelated to the 
classical 
metallicity-dependent weak $s$-component, and might be thought of  as a
lighter element primary process (or LEPP). We stress that the details of this 
nucleosynthesis are still not well understood,  and charged-particle  
reactions and photodisintegrations may contribute along with neutron
production. 
Further, the same process to which 
the light neutron-capture elements Sr, Y and Zr are sensitive, also likely affects 
the production of all elements from Cu to Sr. To understand in 
detail the complicated Galactic nucleosynthesis history of 
Sr, Y and Zr (as well as other lighter element) formation will require 
new theoretical studies and additional high-quality spectroscopic observational data,
particularly of low-metallicity halo stars. 

{\acknowledgements} C.T. thanks the Alexander von Humboldt Foundation, the 
Federal Ministry of Education and Research, and the Programme for Investment 
in the Future (ZIP) of the German Government for their financial support. 
R.G. thanks Max-Planck-Institut f\"ur Cosmochemie (Mainz) for the kind 
hospitality during the development of this work. We thank the 
anonymous referee for helpful suggestions that have improved the paper.  
Research partly supported by the MIUR-FIRB Project
`The astrophysical origin of heavy elements beyond Fe' (CT, RG) and 
by NSF grants AST-9986974  and AST-0307279 (JJC), AST-9987162  
and AST-0307495 (CS).

\newpage
\epsscale{0.85}
\plotone{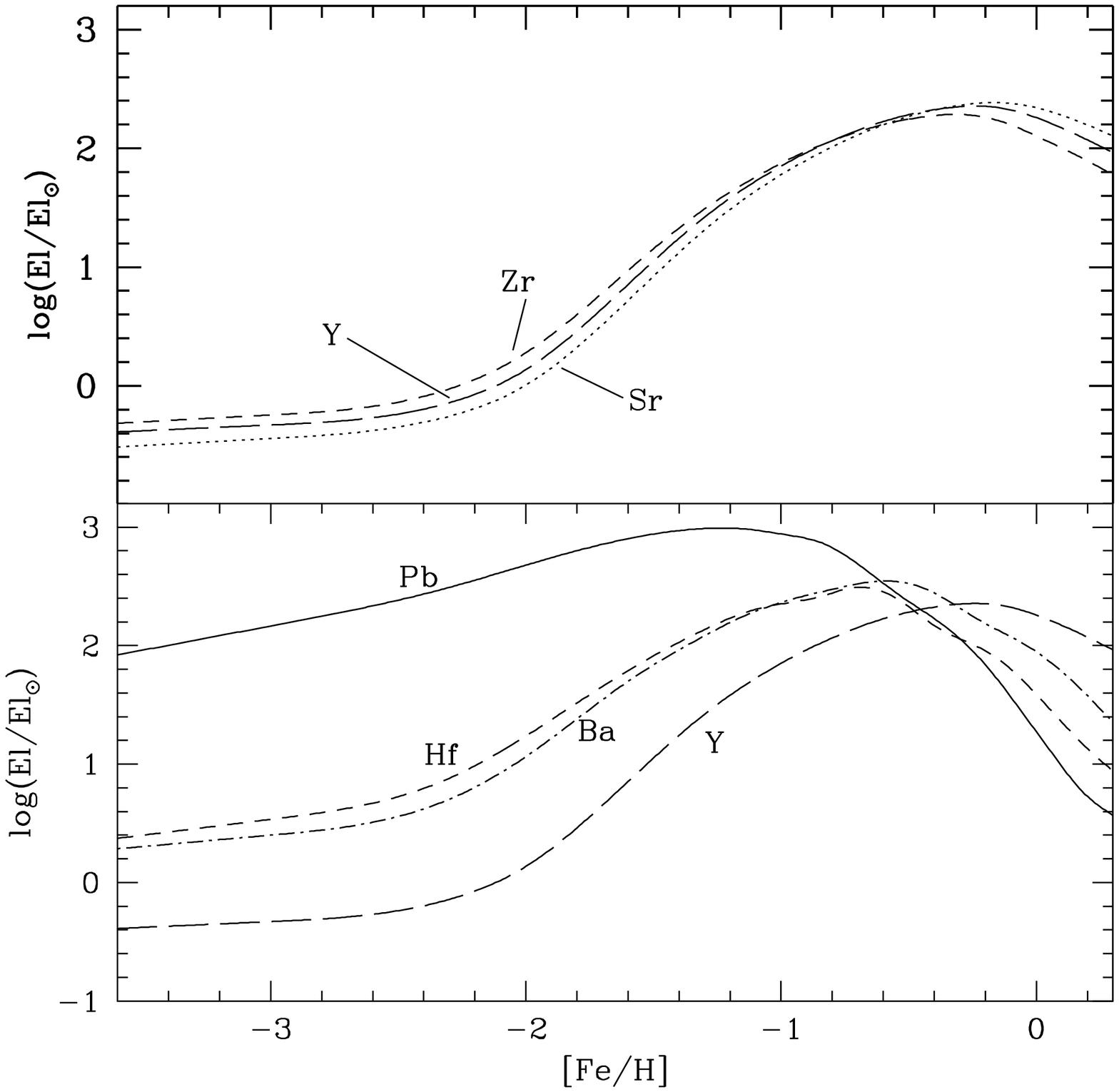}
\figcaption[fig1.eps]{{\small Production factors with respect to the solar 
value in the He intershell material cumulatively mixed by third dredge-up 
episodes with the envelope of AGB stars of 1.5 \msb and 
different metallicities. 
The production of Sr, Y, Zr is shown in the {\it upper panel}. 
In the {\it lower panel} the production of Ba, Hf, and Pb is compared 
with the Y production. The case ST for the $^{13}$C pocket (see text
for details) has been adopted.}
\label{fig1}}

\newpage
\epsscale{0.85}
\plotone{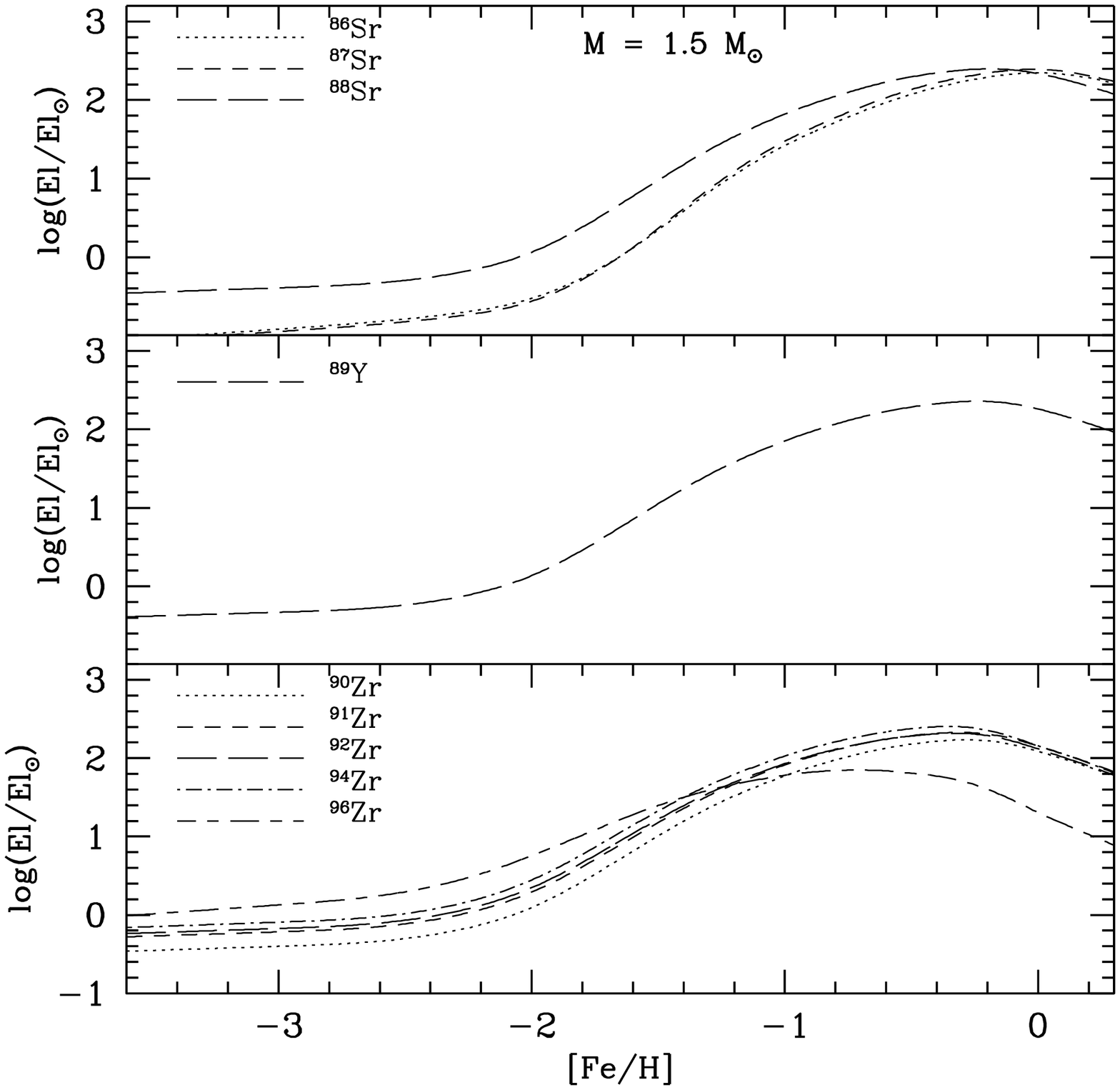}
\figcaption[fig2.eps]{{\small The same of Fig.~1, but for the production of
Sr ({\it upper panel}), Y ({\it middle panel}), and Zr ({\it lower panel}) 
isotopes.}
\label{fig2}}

\newpage
\epsscale{0.85}
\plotone{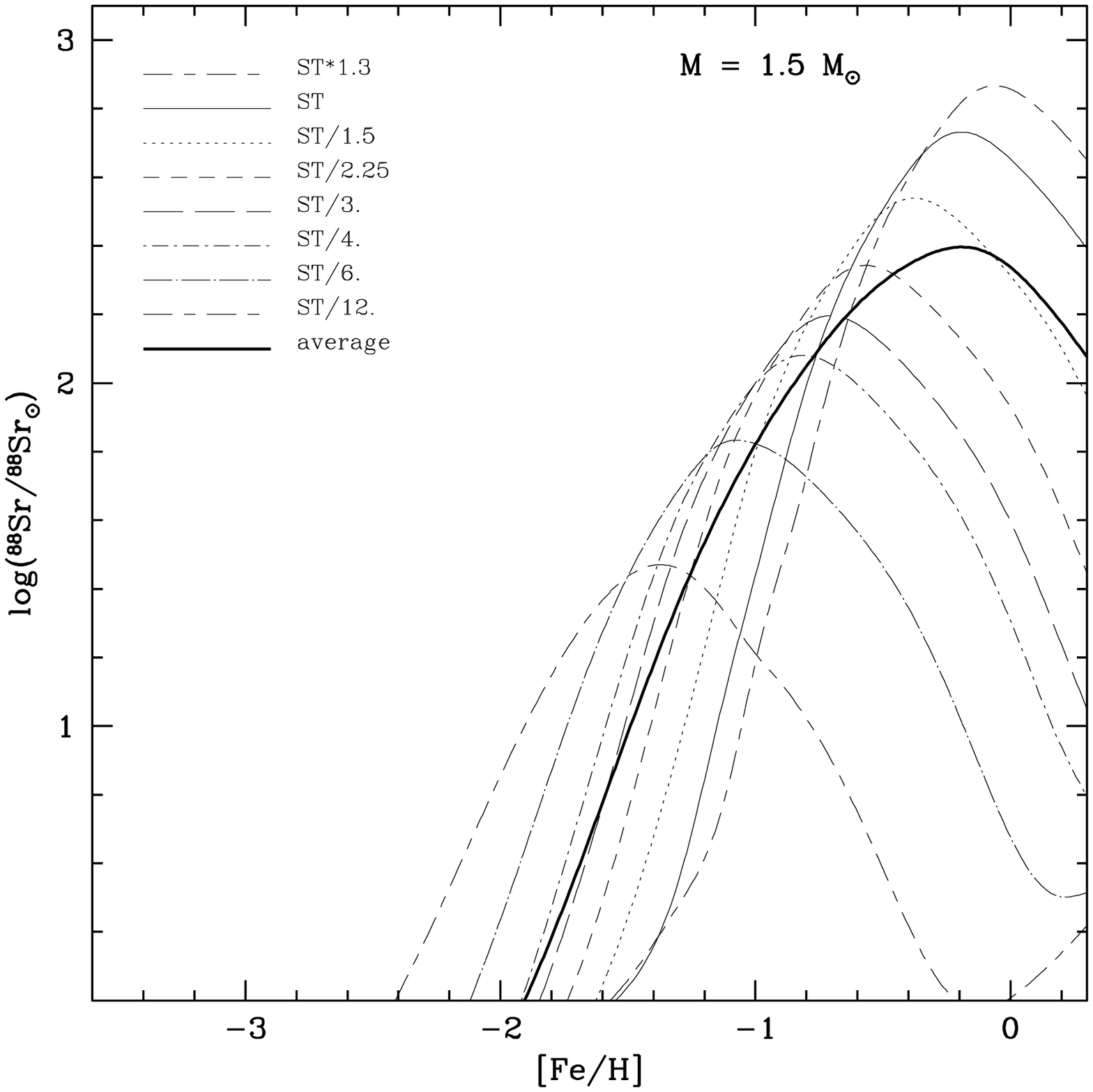}
\figcaption[fig3.eps]{Production factors relative to solar of $^{88}$Sr in 
the He intershell material cumulatively mixed with the envelope of a 
1.5 \msb star by third dredge-up episodes as a function of metallicity, 
for different assumptions on the $^{13}$C concentration in the pocket. 
The {\it thick continuous line} represents the unweighted average of 
all cases shown. 
\label{fig3}}

\newpage
\epsscale{0.85}
\plotone{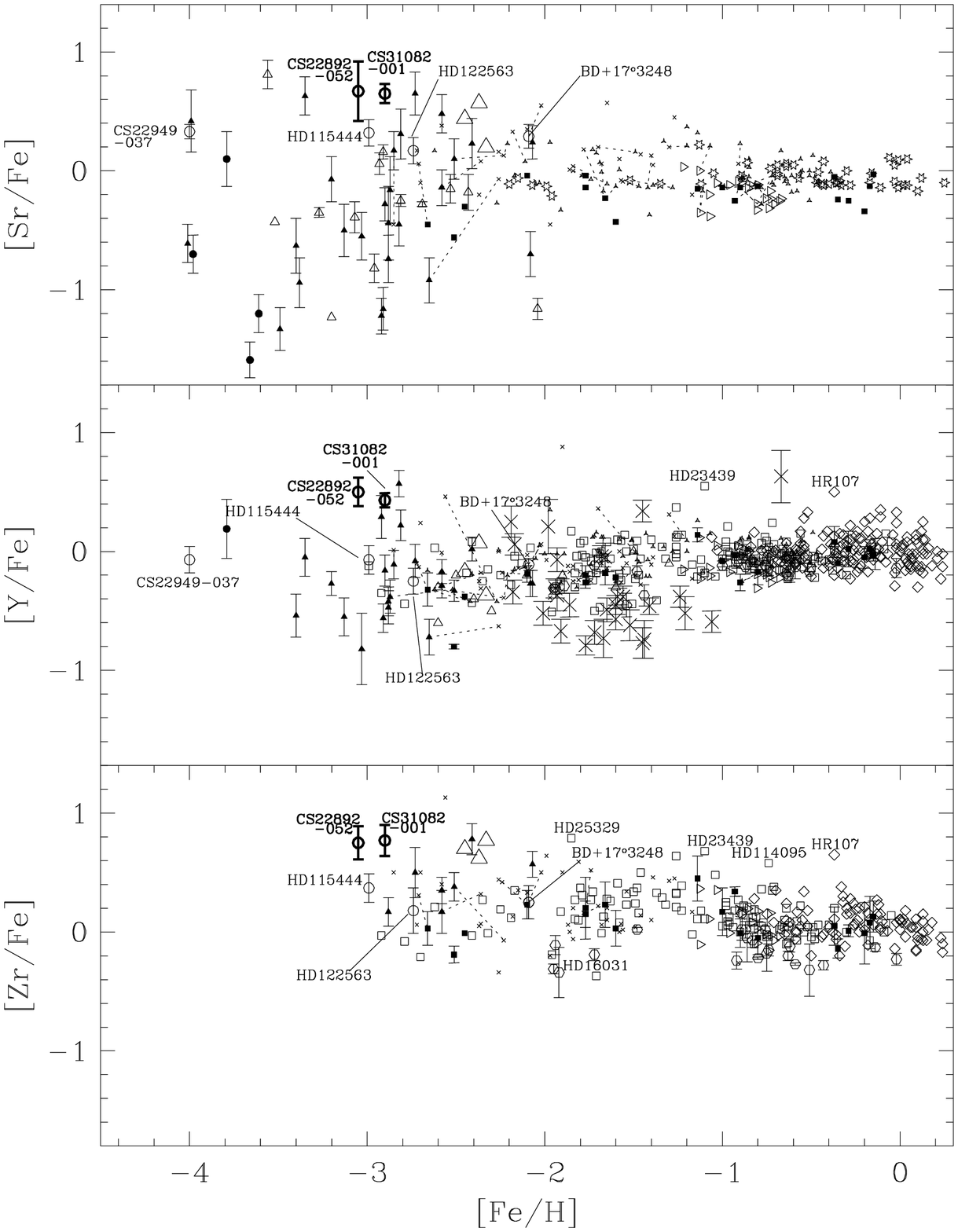}
\figcaption[fig4.eps]{\small{Spectroscopic observations of Galactic disk and 
halo stars at different metallicities for [Sr/Fe] ({\it upper panel}), 
[Y/Fe] ({\it middle panel}), and [Zr/Fe] ({\it lower panel}) from: 
Spite \& Spite~(1978) ({\it open triangles}); Edvardsson et al.~(1993) 
({\it open rhombs}); Gratton \& Sneden~(1994) ({\it filled squares}); 
McWilliam et al.~(1995) and McWilliam~(1998) ({\it filled triangles}); 
Jehin et al.~(1999) ({\it open tilted triangles}); 
Tomkin \& Lambert~(1999) ({\it open hexagons}); Burris et al.~(2000) 
({\it crosses}; Fulbright~(2000) ({\it open squares}); 
Norris, Ryan, \& Beers~(2001) ({\it filled circles}); 
Mashonkina \& Gehren~(2001) ({\it open stars}); Mishenina \& Kovtyukh~(2001) 
({\it small triangles}). 
With {\it open circles} we indicate ``special'' metal-poor stars from:
Westin et al.~(2000), Cowan et al.~(2002), Depagne et al.~(2002).
Two stars, CS22892-052 (Sneden et al.~2000a) and CS31082-001 (Hill et al.~2002),
are indicated as {\it bold open circles} (see text for discussion). 
{\it Big open triangles} are for three stars in M15 (Sneden et al.~2000b, 
and this work). 
{\it Big skeletal} are for stars in dwarf spheroidal galaxies (Shetrone 
et al.~2001,~2003). 
Error bars are shown only when reported for single objects by the authors. 
{\it Thin dotted line} connects a star observed by different authors.
The stars indicated with their names are discussed in detail in the text.} 
\label{fig4}}

\newpage
\epsscale{0.85}
\plotone{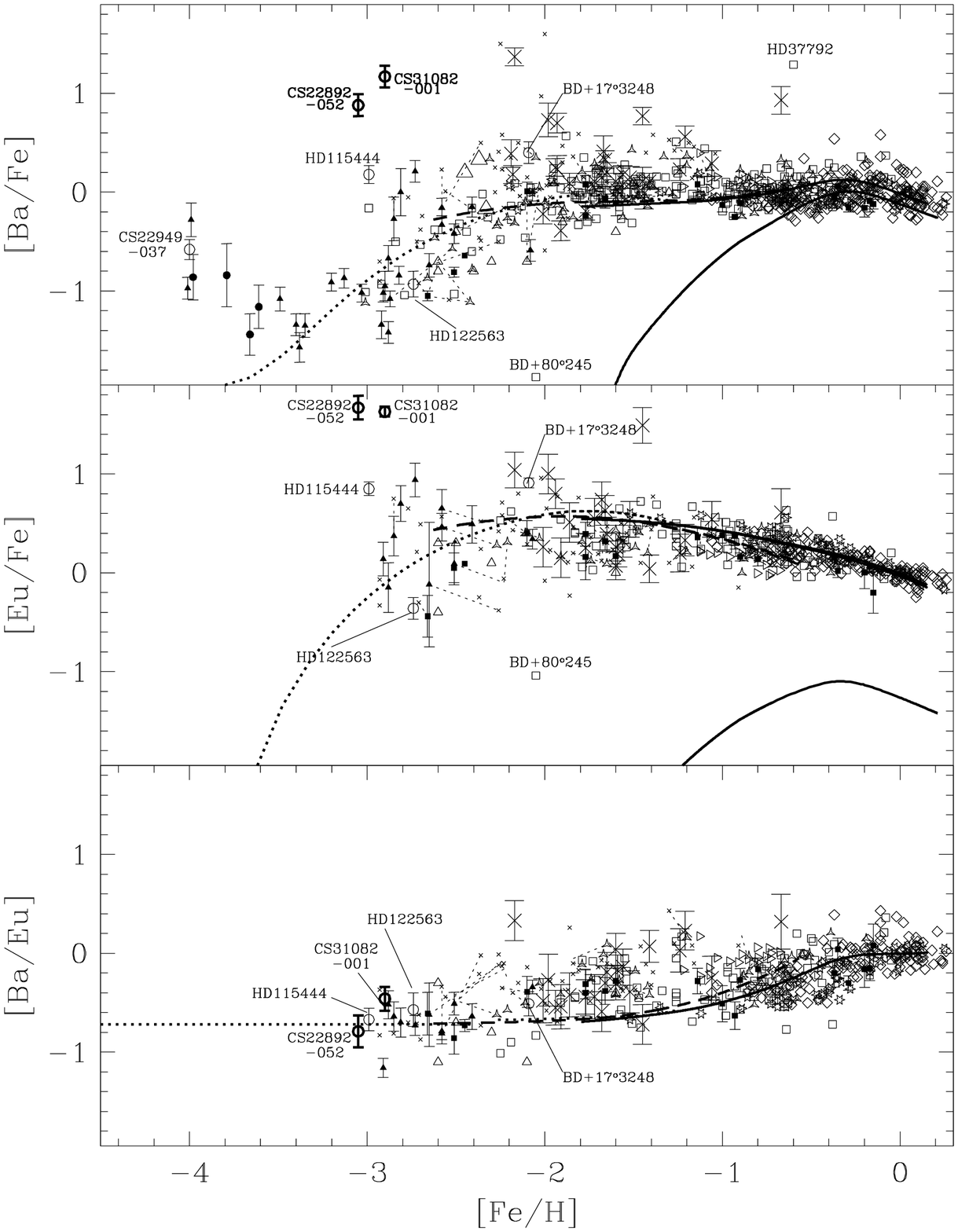}
\figcaption[fig5]{\small{Spectroscopic observations of Galactic disk and
halo stars at different metallicities for [Ba/Fe] ({\it upper panel}),
[Eu/Fe] ({\it middle panel}), and [Ba/Eu] ({\it lower panel}).
The symbols are the same of Fig.~4.
The curves represent the total $s$+$r$ contribution for the halo
({\it dotted lines}), thick-disk ({\it dashed lines}), and thin-disk
({\it solid lines}). With a thin-disk ({\it bold solid line}) we show
for comparison the AGB $s$-process contribution alone.}
\label{fig5}}

\newpage
\epsscale{0.85}
\plotone{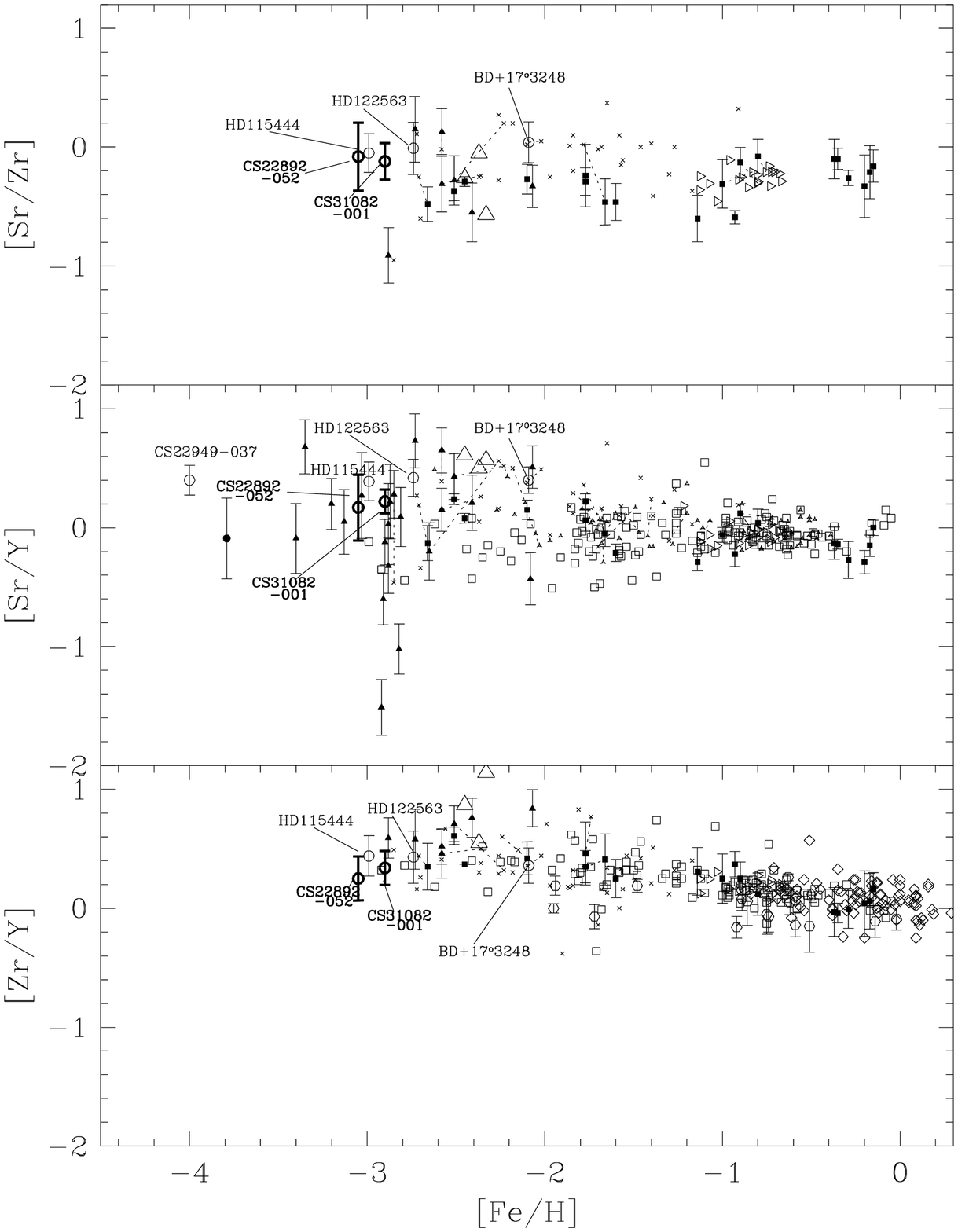}
\figcaption[fig6]{\small{Spectroscopic observations of Galactic disk and 
halo stars at different metallicities for [Sr/Zr] ({\it upper panel}), 
[Sr/Y] ({\it middle panel}), and [Zr/Y] ({\it lower panel}). 
The symbols are the same of Fig.~4. 
The stars indicated with their names are discussed in detail in the text.} 
\label{fig6}}

\newpage
\epsscale{0.85}
\plotone{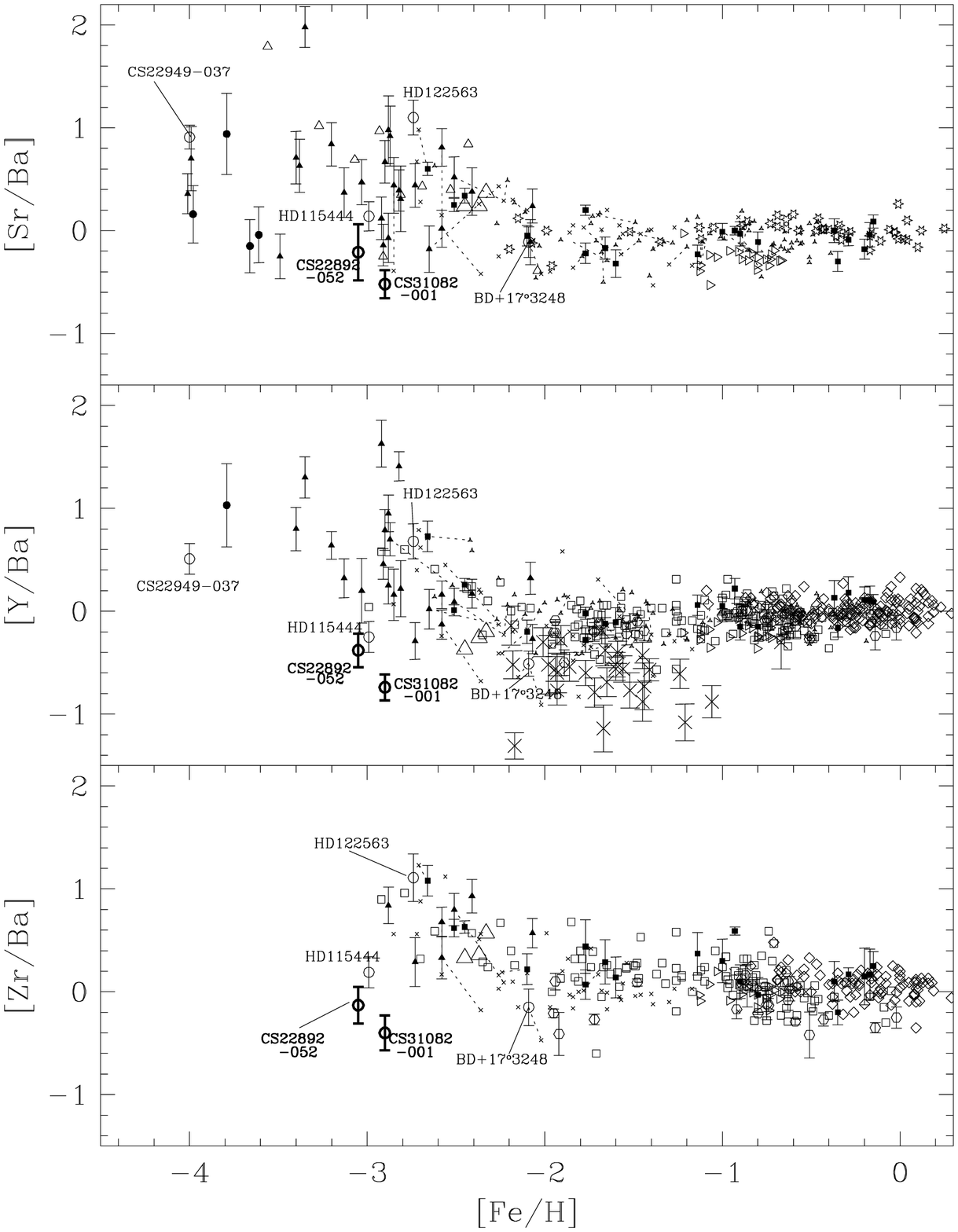}
\figcaption[fig7.eps]{\small{Spectroscopic observations of Galactic disk and 
halo stars at different metallicities for [Sr/Ba] ({\it upper panel}), [Y/Ba] 
({\it middle panel}), and [Zr/Ba] ({\it lower panel}). 
The symbols are the same of Fig.~4. 
The stars indicated with their names are discussed in detail in the text.} 
\label{fig7}}

\newpage
\epsscale{0.85}
\plotone{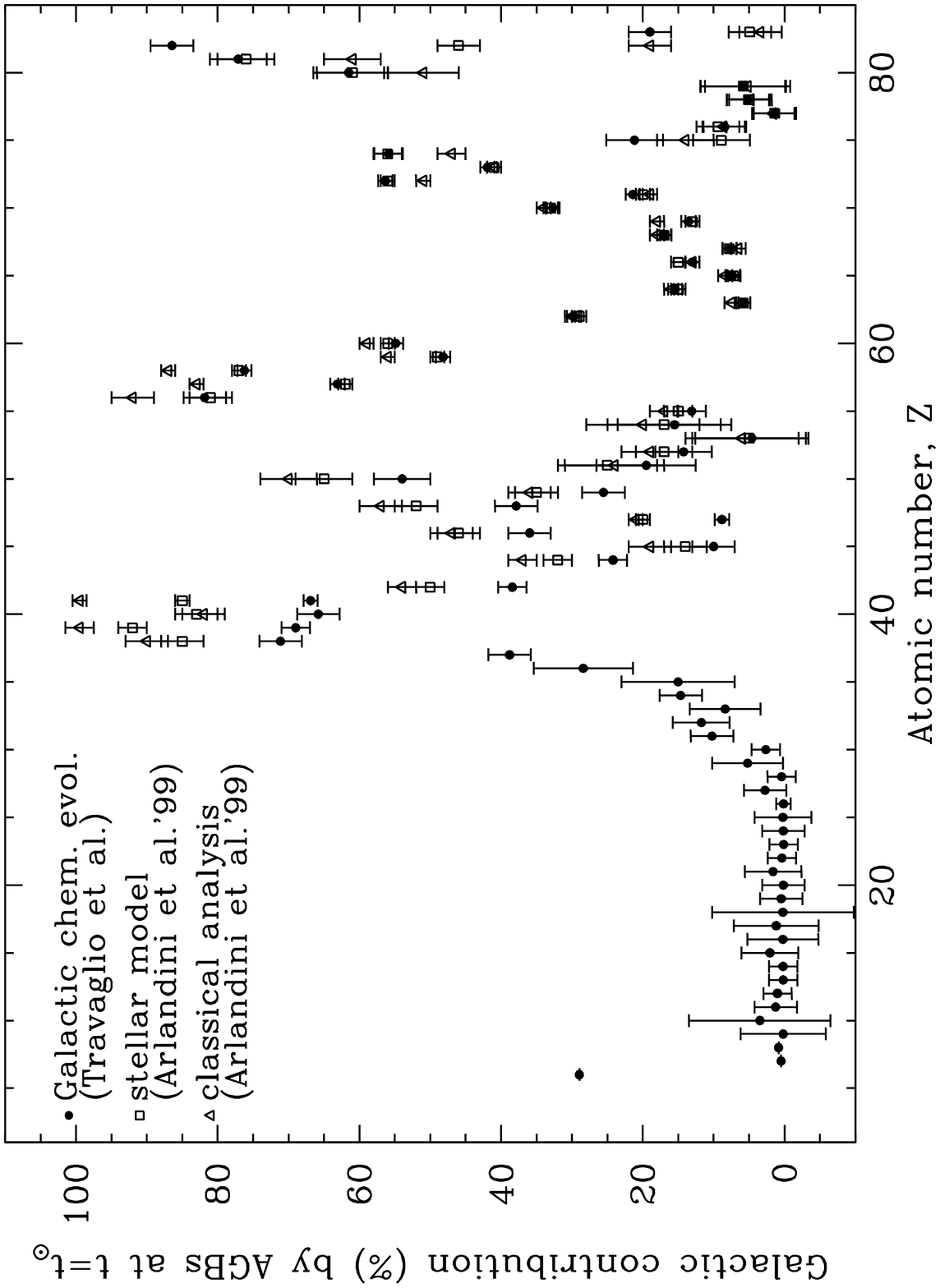}
\figcaption[fig8.eps]{\small{
Galactic contribution by AGB stars at the epoch of the solar system
formation.
Three different models has been considered: the classical analysis 
(Arlandini et al.~1999) ({\it open triangles}), `stellar model'
(Arlandini et al.~1999) ({\it open squares}) and Galactic chemical evolution 
(Travaglio et al.~1999, 2001, this work) ({\it filled dots}).
For  light elements below Fe: 
there is a small $s$-process AGB contribution
to P (2.1\%) and Sc(1.6\%). 
AGBs are responsible for  29\% 
and 3.5\% of the solar carbon and neon, respectively
(Arnone et al. 2003).
For elements beyond Fe and up to Zn, there is a 
minor $s$-process contribution from 
AGBs to  Co (2.7\%), Cu(5.2\%) and Zn(2.6\%).
} 
\label{fig8}}

\newpage
\epsscale{0.85}
\plotone{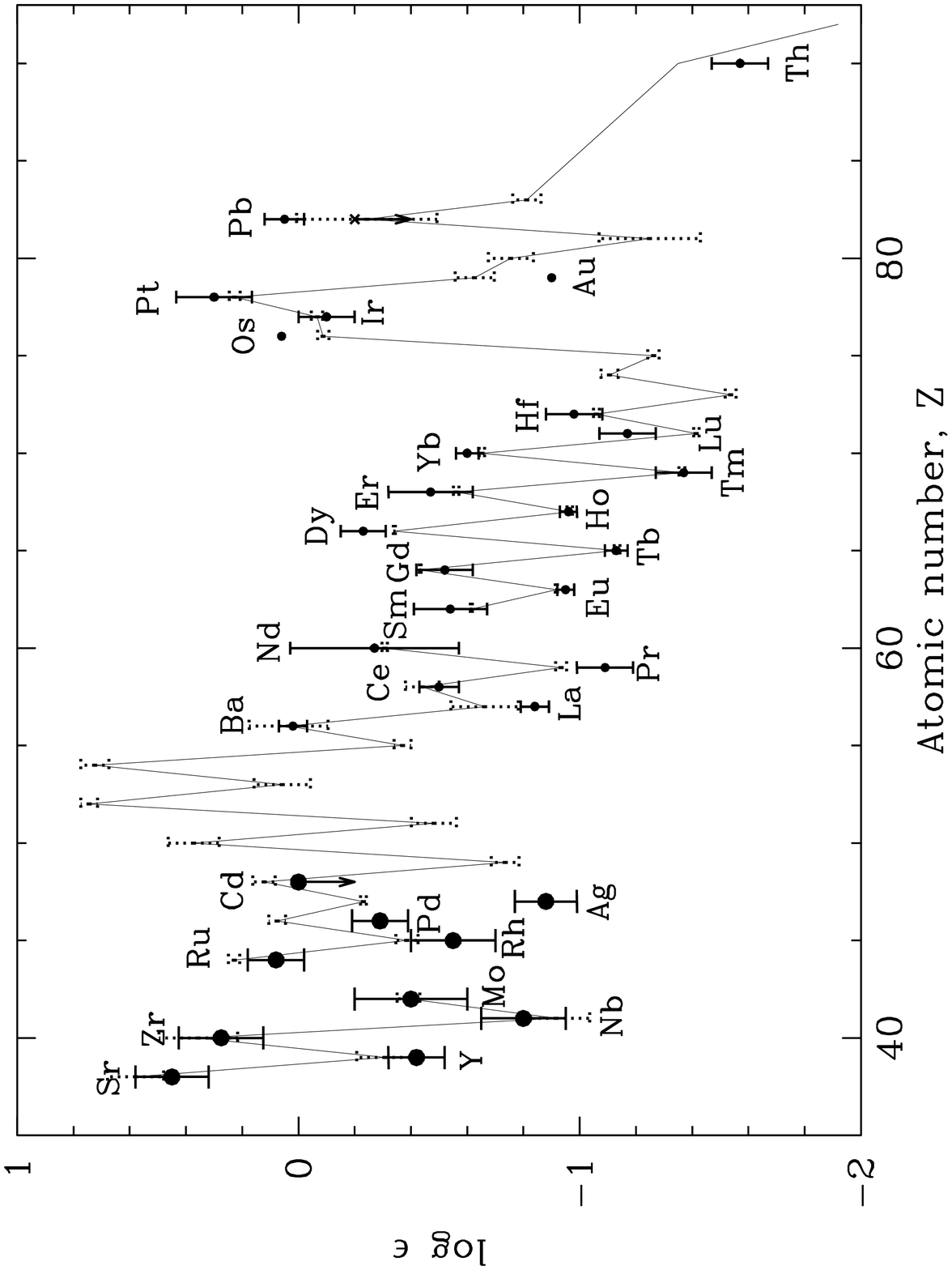}
\figcaption[fig9.eps]{\small{Neutron-capture elements observed in 
CS~22892-052. Ground-based and HST data (see text) are taken into account.
For Pb, the HST measurement (upper limit) has been plotted as a {\it cross}.
Observations are compared to a scaled abundance curve ({\it long-dashed lines}) 
obtained with Galactic chemical evolution calculations described in the text.
Theoretical error bars are also plotted with {\it dotted lines}.} 
\label{fig9}}

\newpage
\epsscale{0.85}
\plotone{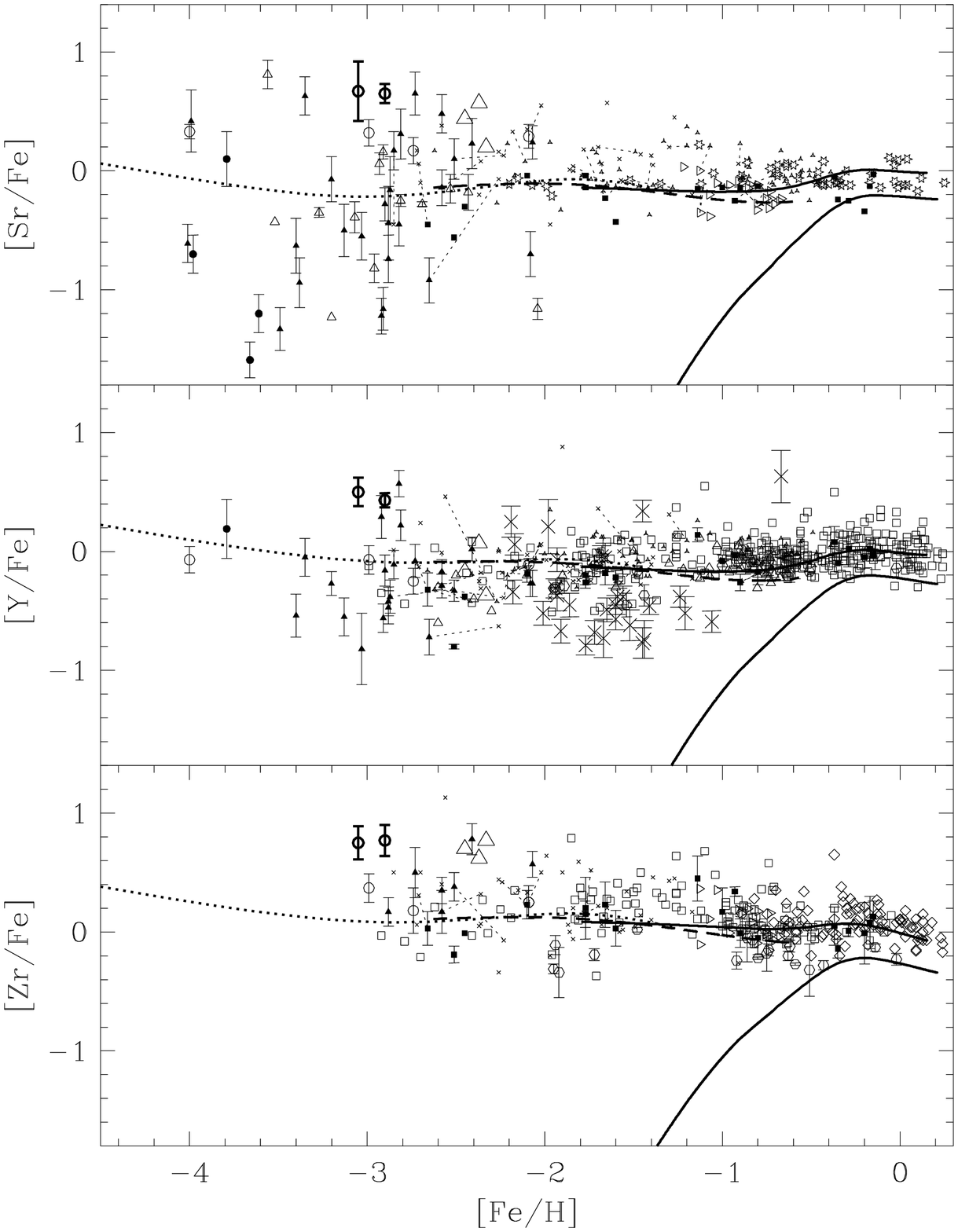}
\figcaption[fig10.eps]{\small{Galactic evolution of [Sr/Fe] 
({\it upper panel}), [Y/Fe] ({\it middle panel}), and [Zr/Fe] 
({\it lower panel}) versus [Fe/H], according to our model predictions 
for the $s$-process, and the total ($s$+$r$+primary-process), 
in the halo ({\it dotted lines}), thick-disk ({\it dashed lines}), 
and thin-disk ({\it solid lines}).
Observational data ({\it open circles}) has been discussed in \S~4 
and shown in Fig.~4.} 
\label{fig10}}

\newpage
\epsscale{0.85}
\plotone{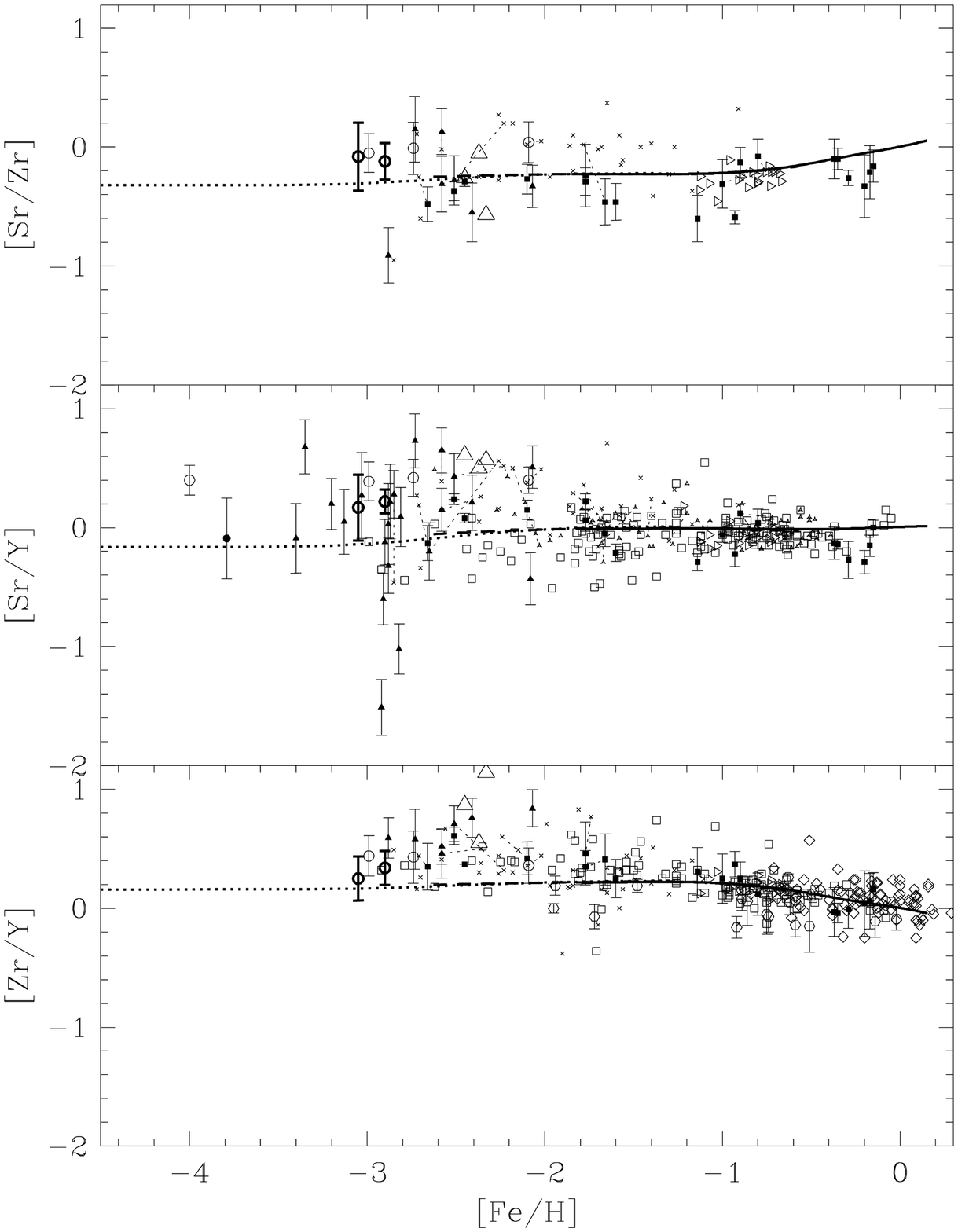}
\figcaption[fig11.eps]{\small{The same of Fig.~10 for [Sr/Zr] 
({\it upper panel}), [Sr/Y] ({\it middle panel}), and [Zr/Y] 
({\it lower panel}) versus [Fe/H]. 
The curves represent the total $s$+$r$+primary contribution for the halo 
({\it dotted lines}), thick-disk ({\it dashed lines}), and thin-disk 
({\it solid lines}).} 
\label{fig11}}

\newpage
\epsscale{0.85}
\plotone{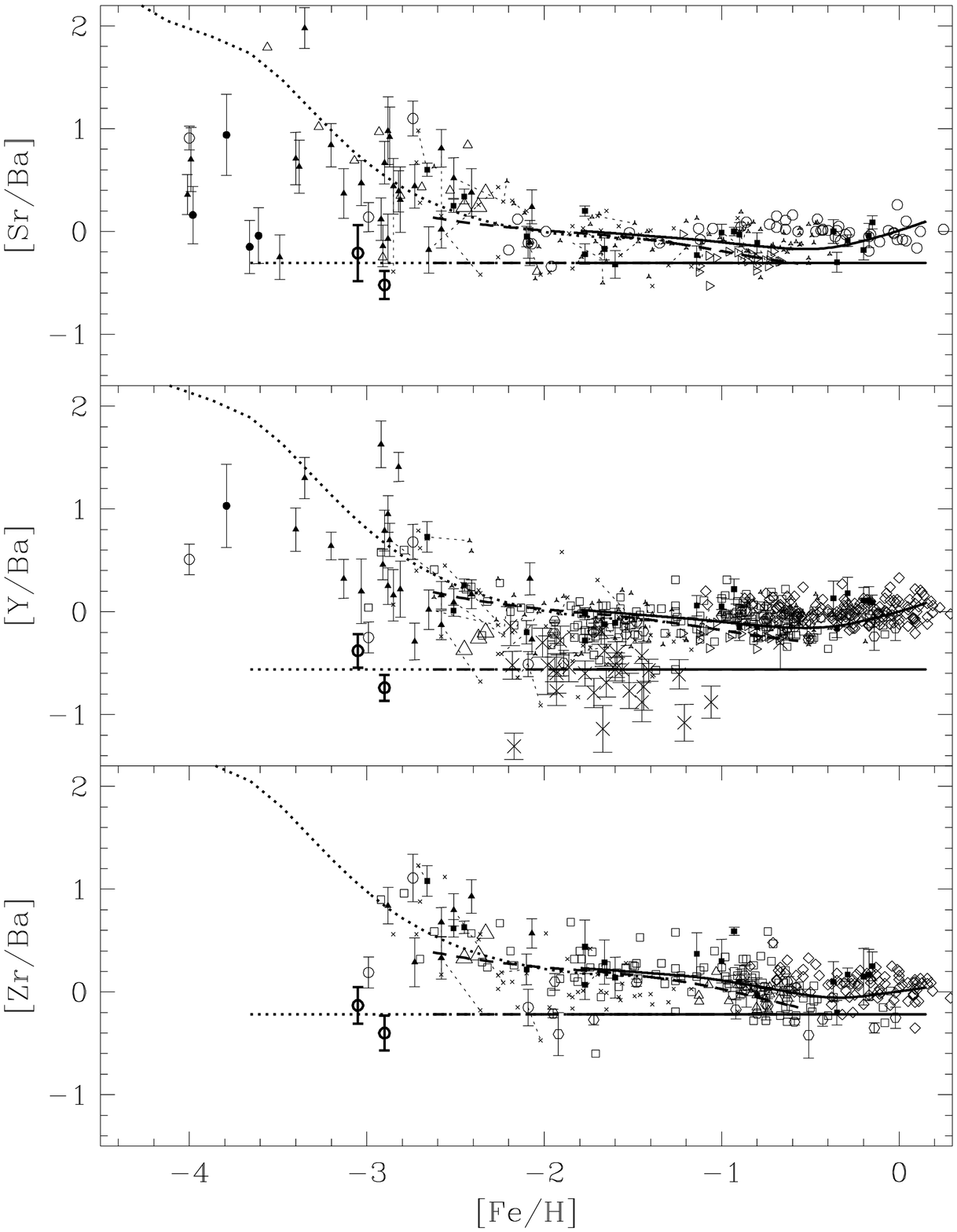}
\figcaption[fig12.eps]{\small{The same of Fig.~10 for [Sr/Ba] 
({\it upper panel}), [Y/Ba] ({\it middle panel}), and [Zr/Ba] 
({\it lower panel}) versus [Fe/H] for the halo ({\it dotted 
lines}), thick-disk ({\it dashed lines}), and thin-disk 
({\it solid lines}).With {\it thick lines} the $r$-process contribution
alone is also plotted for comparison.}
\label{fig12}}

\end{document}